\documentclass[journal=jacsat,manuscript=article]{achemso}

\usepackage{xcolor}
\usepackage{amsmath}
\usepackage{siunitx}

\newcommand{\handlefoot}{\footnote{Note that other experimental and theoretical studies have determined one-photon ionization thresholds to be \SI{2.64}{eV} (\SI{470}{nm})\cite{Deak2014}, \SI{2.7}{eV} (\SI{459}{nm})\cite{Bourgeois2017} and \SI{2.74}{eV} (\SI{452}{nm})\cite{Bourgeois2017} instead which are in good agreement with the value of \SI{2.6}{eV} (\SI{477}{nm}) of Aslam \textit{et al.}\cite{Aslam2013} that we utilize here. The threshold for one-photon recombination has not yet been experimentally measured, but was calculated here by subtracting the energy difference between the ground state of NV$^-$ and the conduction band from the bandgap of diamond. }}

\author{Laura A. Völker}
\affiliation{Department of Physics, ETH Zürich, Otto-Stern-Weg 1, 8093 Zürich, Switzerland}
\author{Konstantin Herb}
\affiliation{Department of Physics, ETH Zürich, Otto-Stern-Weg 1, 8093 Zürich, Switzerland}
\author{Darin A. Merchant}
\affiliation{Department of Physics, ETH Zürich, Otto-Stern-Weg 1, 8093 Zürich, Switzerland}
\author{Lorenzo Bechelli}
\affiliation{Department of Physics, ETH Zürich, Otto-Stern-Weg 1, 8093 Zürich, Switzerland}
\author{Christian L. Degen}
\affiliation{Department of Physics, ETH Zürich, Otto-Stern-Weg 1, 8093 Zürich, Switzerland}
\altaffiliation{Quantum Center, ETH Zürich, 8093 Zürich, Switzerland}
\email{degenc@phys.ethz.ch}
\author{John M. Abendroth}
\email{jabendroth@phys.ethz.ch}
\affiliation{Department of Physics, ETH Zürich, Otto-Stern-Weg 1, 8093 Zürich, Switzerland}

\title{Charge and Spin Dynamics and Destabilization of Shallow Nitrogen--Vacancy Centers under UV and Blue Excitation}

\begin{document}

\begin{center}
	\includegraphics[width=0.7\textwidth]{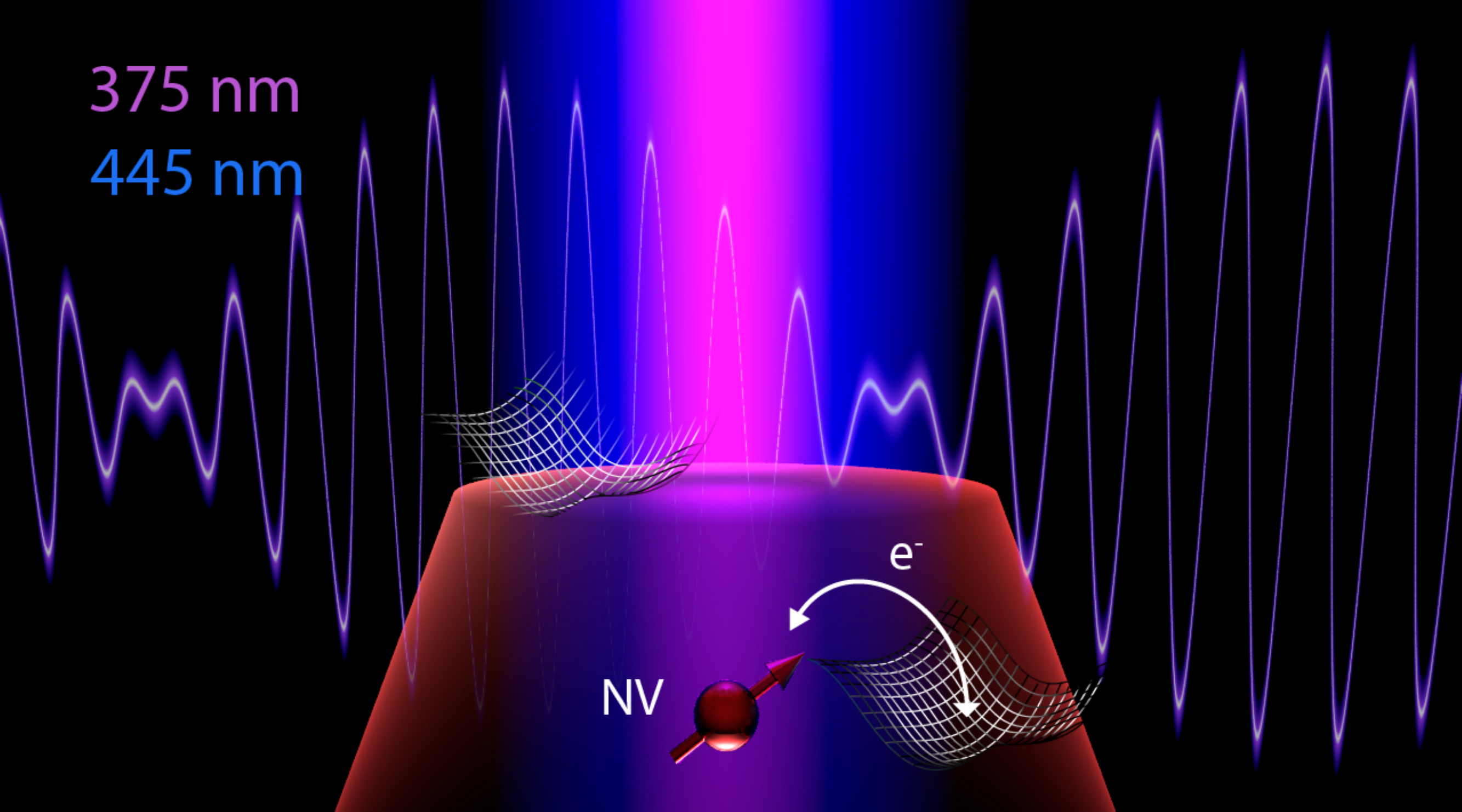}
\end{center}

\newpage
\section{Abstract}

Shallow nitrogen--vacancy (NV) centers in diamond offer unique opportunities for studying photochemical reactions at the single-molecule level, such as the photogeneration of radical pairs in proximal molecules. A prerequesite for such experimental schemes is the detailed understanding of the charge and spin dynamics of NV centers exposed to the short-wavelength light required for photoexciting chemical species. Here, we measure and analyze the charge and spin dynamics of shallow NV centers under \SI{445}{nm} (blue) and \SI{375}{nm} (UV) illumination. With blue excitation, we observe a power-dependent charge-state preparation accompanied by modest preservation of spin initialization fidelity. Under UV excitation, we find a power-independent charge-state preparation and no spin polarization. We further observe an irreversible aging of NV centers under prolonged exposure to UV, and to a lesser extent, blue laser excitation, which we attribute to formation of new electronic trap states. This aging manifests itself in a reduced charge stability and spin contrast, and is detrimental to the NV sensing performance. We evaluate the prospects and limitations of NV centers for probing photogenerated radical pairs based on experimentally measured sensitivities following blue and UV excitation, and outline the design rules for possible sensing schemes. 

\section{Keywords}
\textit{nitrogen--vacancy (NV) center, quantum sensing, photoionization, optically detected magnetic resonance (ODMR), radical pairs}\newline

\newpage
Photoexcitation by blue and UV light plays a pivotal role in biology, where illumination initiates chemical transformations, often \textit{via} formation of short-lived radicals and radical pairs (RPs).\cite{Kothe1994, Levanon1997,Schmermund2019, Hore2016, Biskup2009, Hadi2022} Further, photogenerated RPs in synthetic chemical architectures excited by these wavelengths are of interest as molecular qubits for quantum information processing.\cite{Mani2022, Rugg2019, Nelson2020, Mao2023} To address these systems \textit{via} magnetic resonance spectroscopy in the single- to few-molecule regime, quantum sensing with near-surface nitrogen--vacancy (NV) centers in diamond has been proposed to track charge separation and recombination, spin dynamics, and spin polarization in photogenerated RPs.\cite{Liu2017, Finkler2021, Voelker2023,Khurana2024} The NV exhibits high magnetic field sensitivity, nanoscale localization, and room-temperature operability, offering complementary information to conventional EPR and NMR spectroscopies.\cite{Lovchinsky_science_2016,Janitz2022,Abendroth2022,LiuBucher2022,Xie2022} Thus, magnetic field sensing with shallow NVs could provide insight into the complex spin dynamics and photochemical cascades involved in RP generation.

Experimental realization of NV-RP sensing requires a detailed understanding of charge ionization rates, spin dynamics, and photostability of shallow defects under blue or UV illumination.\cite{Rodgers2024,Martinez2020,Ninio2021} Despite a significant body of work on NV center photoionization over a wide range of excitation wavelengths, \cite{Gaebel2006,Waldherr2011, Beha2012, Aslam2013,Hacquebard2018, Razinkovas2021} effects of blue and UV irradiation are little characterized by comparison. Previous works that explored the influence of irradiation in the near-UV and blue regions utilized all-optical methods to track interconversion between the desired negatively charged NV$^-$ and undesired neutral NV$^0$ states.\cite{Han2010, Aslam2013, Chen2013,Bourgeois2017,Li2022, Yang2022, Wood2024} Still, detailed effects on NV spin dynamics, such as the spin initialization fidelity, remain to be explored in detail. Importantly, exposure to short wavelengths may also accelerate deterioration of shallow NV properties, and necessitates further analysis. Such effects are evidenced by, \textit{e.g.}, blinking, loss in optically detected magnetic resonance (ODMR) contrast, and accelerated ionization rates due to changes to the NV's electronic environment.\cite{Bradac2010,Dhomkar2018, Sangtawesin2019, Bluvstein2019, Yuan2020}

In this work, we expand on previous studies of blue and UV photophysics of diamond defects \cite{Han2010, Aslam2013, Chen2013,Bourgeois2017,Li2022, Yang2022, Wood2024} by investigating the effects of \SI{445}{nm} and \SI{375}{nm} excitation on the charge and spin dynamics of individual, shallow (\textit{ca}. 10-nm-deep) NV centers. Under \SI{445}{nm} laser excitation, we observe power-dependent charge-state initialization, rationalized by one-photon ionization and two-photon recombination dynamics. Decent spin-state initialization fidelities are maintained at ca. 50\% of the original ODMR contrast under steady-state conditions. Under UV excitation, we observe power-independent charge-state initialization and complete absence of optical spin polarization. We further find that prolonged exposure to either wavelength permanently alters the NV ionization dynamics. These effects are attributed to the formation of modified electronic environments, likely \textit{via} generation of additional electron traps. Finally, we outline the prospects of shallow defects for NV-RP sensing under realistic experimental conditions.

\begin{figure}
\includegraphics[width=1.0\textwidth]{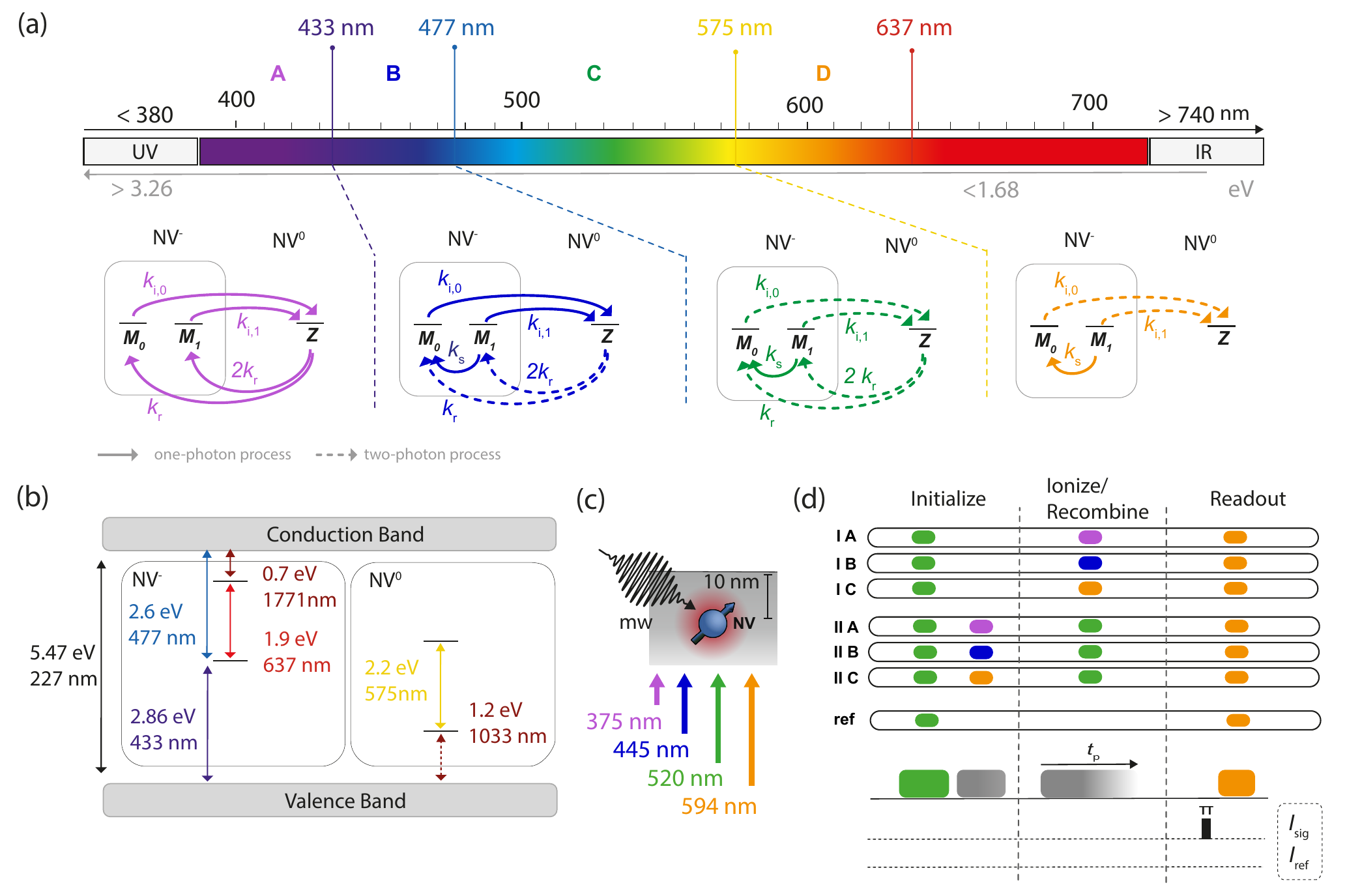}
\caption{\textbf{Optically-driven charge-state dynamics of the NV center and experimental protocol.} \textbf{(a)} \textit{Top}: The relevant wavelength ranges can be divided into four regions, A-D, depending on the ionization and recombination pathways which are accessible at a given wavelength. \textit{Bottom}: Schematic depiction of a simplified three-level system, modeling NV$^{-}$ with two levels ($M_0$ and $M_1$) for spin states $m_S = 0$ and $m_S = \pm1$, respectively and NV$^{0}$ as a single level ($Z$). Spin-dependent ionization rates are denoted as $k_{\mathrm{i},0}$ and $k_{\mathrm{i},1}$, $k_{\mathrm{s}}$ is the rate of spin-state polarization, and $k_{\mathrm{r}}$ is the recombination rate from NV$^{0}$ to NV$^{-}$. Solid lines represent one-photon excitations, dashed lines indicate two-photon excitations.   \textbf{(b)} Energy level diagram of the electronic ground and excited states of NV$^{-}$ and NV$^{0}$ within the band gap of the diamond host material. \textbf{(c)} Schematic of the experimental set-up: A diamond membrane hosting single, shallow NV centers is exposed to different laser colors in a confocal microscope with integrated microwave infrastructure. \textbf{(d)}  Experimental protcol, \textit{i.e.} pulse sequences utilized to study charge and spin state manipulation of shallow NV centers.}
\label{revfig1}
\end{figure}

\textbf{Figure \ref{revfig1}a,b} illustrates the photoionization dynamics between NV$^{-}$, which exhibits a zero phonon line (ZPL) at \SI{637}{nm}, and NV$^{0}$ with a ZPL at \SI{575}{nm}.\cite{Gaebel2006} Interconversion between the charge states reduces the ODMR  contrast by decreasing spin-state initialization fidelity of NV$^-$ and increasing fluorescence background due to a high steady-state NV$^0$ population.\cite{Bluvstein2019}  The essential dynamics of the system are captured by a three-level model (\textbf{Figure \ref{revfig1}a,} and SI) that comprises two levels ($M_0$ and $M_{1}$) for NV$^-$ in the $m_S = 0$ and $m_S = \pm1$ states, respectively, and one level ($Z$) for NV$^0$.  Charge and spin dynamics are described by spin-dependent ionization rates $k_{\mathrm{i},0}$ and $k_{\mathrm{i},1}$, spin-state polarization with rate $k_{\mathrm{s}}$, and charge recombination at rate $k_{\mathrm{r}}$. The three-level model forms the basis for distinguishing four dynamic regimes in the UVA-visible spectrum, categorized by one- \textit{versus} two-photon ionization or recombination of NV charge states. At wavelengths between the ZPL of NV$^0$ and NV$^-$  (region D), photons primarily excite NV$^-$ but not NV$^0$. This selective excitation of NV$^-$ enables charge-state-sensitive protocols.\cite{Shields2015, Hopper2018} However, high laser powers trap the population in NV$^0$ due to two-photon ionization, and re-ionization of NV$^-$ requires irradiation with shorter wavelengths.\cite{Han2010} Therefore, sensing with NVs commonly utilizes wavelengths $\sim\SI{532}{nm}$ (region C, \SI{476}{nm}--\SI{575}{nm}), where both charge states are excited and interconverted \textit{via} two-photon ionization and recombination allowing for NV$^{-}$ populations of \textit{ca.} 70\%.\cite{Waldherr2011, Aslam2013} In addition, optical spin polarization (initialization) can be achieved with high fidelity, especially at low laser powers.\cite{Ernst2023}   At wavelengths shorter than \SI{477}{nm}, but longer than \SI{433}{nm} (region B), optical spin initialization competes with the onset of one-photon ionization of NV$^-$ \textit{via} loss of an electron to the conduction band,\cite{Aslam2013, Razinkovas2021} while recombination of NV$^0$ to NV$^-$ remains a two-photon process.\handlefoot  Only at wavelengths below \SI{433}{nm} (region A) are one-photon pathways accessible for both ionization and recombination. Although the energy levels that dictate these thresholds are well-established,\cite{Razinkovas2021} experimental studies have categorized wavelengths below \textit{ca.} \SI{450}{nm} as a uniform regime with charge-state dynamics that are dominated by one-photon ionization of NV$^-$ from the ground state.\cite{Chen2013,Bourgeois2017,Wood2024} In the following, we demonstrate that regions A and B feature distinct photodynamics, especially at higher laser powers. Further, we show that loss of spin-initialization fidelity at these wavelengths is due to a combination of reduced photoluminescence excitation (PLE)\cite{Beha2012} and one-photon ionization processes.

Our measurement scheme and available laser wavelengths are depicted in \textbf{Figure \ref{revfig1}c,d}. Experiments are performed on a commercially available (QZabre Ltd) (001)-terminated single-crystalline CVD-grown diamond membrane implanted with $^{15}$N$^{+}$ ions at 7 keV energy and featuring nanofabricated pillar waveguide arrays to enhance optical collection efficiency and register single NVs.\cite{Zhu2023} The membrane sample is annealed in O$_2$ at 460 \textdegree C before characterization to remove graphitic carbon and terminate the surface with oxygen.\cite{Sangtawesin2019,Abendroth2022} Experimental protocols are schematically outlined in \textbf{Figure \ref{revfig1}d}. First, we probe photoionization dynamics and spin state perturbation under \SI{375}{nm} or \SI{445}{nm} excitation with comparison to orange (\SI{594}{nm}) illumination as a common reference (IA--IC, \textbf{Figure \ref{revfig1}d}). In each protocol, charge and spin state of NVs are first initialized using a green (\SI{520}{nm}) laser; then a perturbing \SI{375}{nm}, \SI{445}{nm}, or \SI{594}{nm} laser pulse of varying length $t_p$ and power $P$ is applied. Second, we analyze re-initialization of the NV from the perturbed states, \textit{i.e.}, the recombination dynamics (IIA--IIC, \textbf{Figure \ref{revfig1}d}).  Here, the NV is first prepared in its equilibrium charge and spin state under  \SI{375}{nm}, \SI{445}{nm}, or \SI{594}{nm} illumination before being re-initialized with a green laser pulse of length $t_p$. In all experiments, the resulting state of the NV is read out in a pulsed ODMR experiment. To manipulate the spin state, we apply resonant microwave pulses supplied by a coplanar waveguide and apply a 10 mT magnetic bias field oriented along the NV axis to lift the $m_s = \pm 1$ degeneracy. For readout, we use a low-power orange pulse enabling simultaneous detection of the NV spin and charge states. \cite{Shields2015, Hopper2018}  All relevant laser pulse durations, powers and delay times for data presented in the main text are detailed in the SI.

Our two experimental measurables are the population of the NV$^-$ charge state $\rho$  and the ODMR contrast $c$ describing the degree of spin polarization in NV$^-$. In the context of our three-level model, we define $\rho$ and $c$ by 
\begin{subequations}
\begin{equation}
\rho = \frac{M_0 + M_1}{ M_0 + M_1 + Z}
\label{eq:rhomodel}
\end{equation}

\begin{equation}
c = \frac{M_0-M_1}{M_0},
\label{eq:cmodel}
\end{equation}
\end{subequations}
with level populations $M_0$, $M_1$ and $Z$. Under orange laser excitation, only populations $M_0$ and $M_1$ contribute to the fluorescence signal. Thus, our experimental observables are directly proportional to $M_0$ and $M_1$. Following a short derivation (see SI), we approximate $\rho$ and $c$ for each laser pulse duration by 

\begin{subequations}
\begin{equation}
\rho(t_p) \approx \frac{\frac{1}{3}I_{\mathrm{ref}}(t_p) + \frac{2}{3}I_{\mathrm{sig}}(t_p) }{\frac{1}{3}I_{\mathrm{ref}}^0(t_p) +\frac{2}{3}I_{\mathrm{sig}}^0(t_p)}
\label{eq:rhoexp}
\end{equation}

\begin{equation}
c(t_p) \approx \frac{I_{\mathrm{ref}}(t_p)-I_{\mathrm{sig}}(t_p)}{I_{\mathrm{ref}}(t_p)}
\label{eq:cexp}
\end{equation}
\end{subequations}
where $I_{\mathrm{sig}}(t_p)$ and $I_{\mathrm{ref}}(t_p)$ are the photoluminescence intensities in signal and reference traces of the ODMR experiment. 
Since the absolute value of $\rho$ cannot be determined experimentally, we calculate $\rho$ relative to the charge state under green laser illumination by normalizing with respect to ODMR signals $I_{\mathrm{ref}}^0$ and  $I_{\mathrm{sig}}^0$ obtained in a separate reference measurement (ref, \textbf{Figure \ref{revfig1}d}). From  $\rho(t_p)$  we extract ionization rates $k_i^{\lambda}$ ($\lambda$=\SI{375}{nm}, \SI{445}{nm}, \SI{594}{nm}) for protocols IA--IC and recombination rates $k_r^{520}$ for protocols IIA--IIC. The equilibrium charge and spin states  $\rho^{\lambda}$ and $c^{\lambda}$ are obtained at long $t_p$, \textit{i.e.}, $(t_p \rightarrow \infty)$ under a given illumination wavelength $\lambda$.

 \begin{figure}
\includegraphics[width=1.0\textwidth]{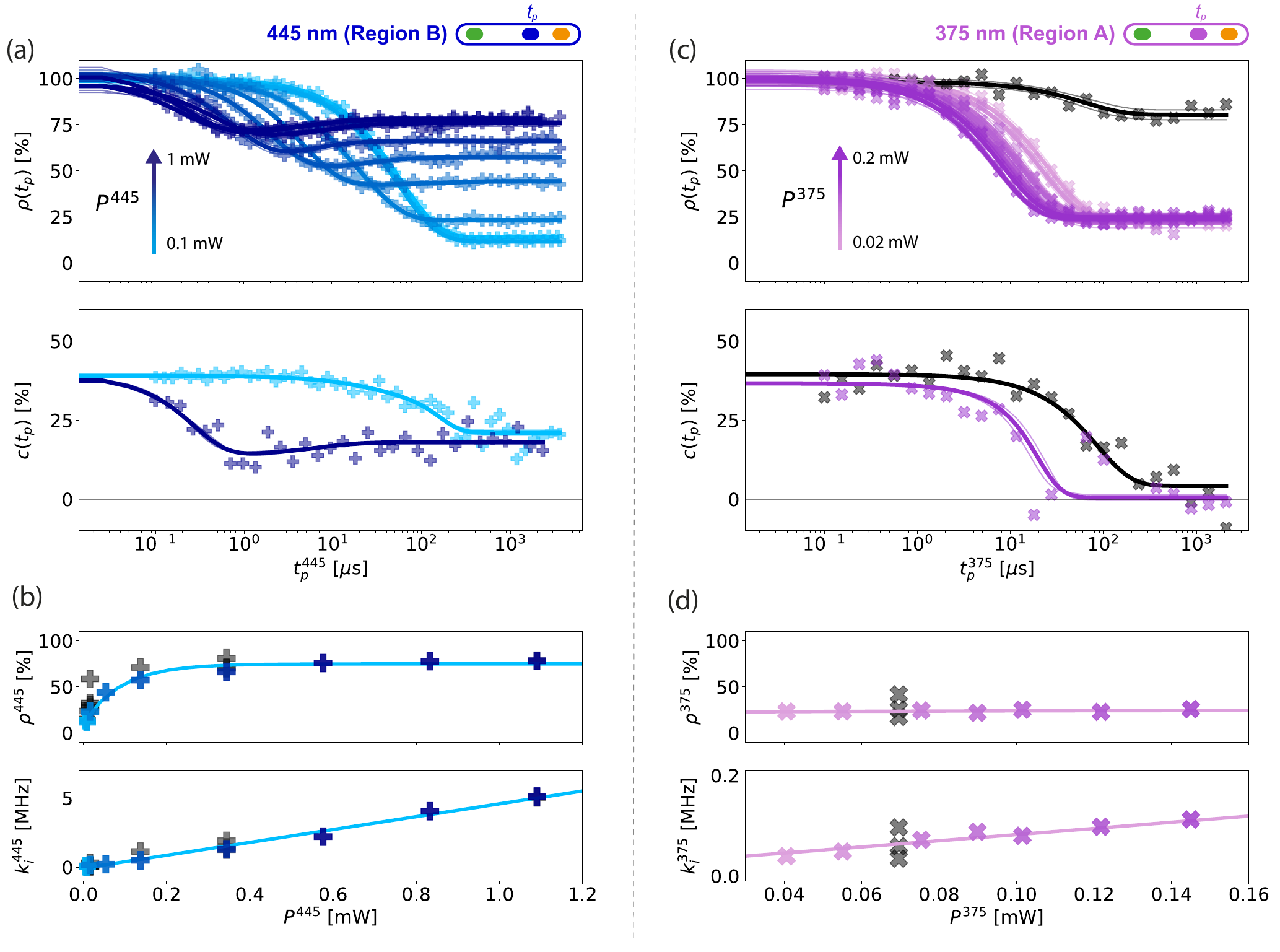}
\caption{\textbf{Charge and spin dynamics  under 445 nm illumination (\textit{left}) and 375 nm illumination (\textit{right}) for representative NVs.}  \textbf{(a,c)} Evolution of NV$^-$ population $\rho$ (\textit{Top}) and ODMR contrast $c$ (\textit{Bottom}) under \SI{445}{nm} (\SI{375}{nm}) laser illumination at different laser powers, $P^{445}$ ($P^{375}$), encoded in marker color. Fitting was performed on the raw experimental data, $I_{\mathrm{ref}}(t_p), I_{\mathrm{sig}}(t_p)$, from which fits for $\rho$ and $c$ were calculated (Equations \ref{eq:rhoexp}, \ref{eq:cexp}). \textbf{(b,d)} Power dependence of charge and spin dynamics under  \SI{445}{nm} (\SI{375}{nm}) illumination, characterized \textit{via} NV$^-$ population $\rho^{445}$ ($\rho^{375}$) and ionization rate $k_i^{445}$ ($k_i^{375}$). Initial measurements that were performed before the NV had sufficiently stabilized in an aged state are shown in grey. An especially pronounced change is evident for UV charge dynamics, where pristine NV centers ionize to a much lesser extent despite complete loss of ODMR contrast. }
\label{revfig2}
\end{figure}

We begin our study by investigating the charge and spin dynamics in wavelength regime B. \textbf{Figure \ref{revfig2}a,b} shows data for a representative NV subjected to \SI{445}{nm} excitation between 0.1 and 1 mW (protocol IB). At low powers ($P^{445} \sim$ \SI{0.1}{mW}) significant ionization to $\rho^{445}\leq$ 20\% occurs, aligning with previous work.\cite{Chen2013,Aslam2013,Bourgeois2017} Yet, $\rho^{445}$ increases up to $\geq$75\%  if $P^{445}$ is raised to $\sim$\SI{1}{mW}. This observation is in agreement with our preceding theoretical considerations for regime B. Here, the energy of a single photon is sufficient for ionization, while charge recombination requires two photons. The resulting  linear \textit{versus} quadratic power dependencies of  ionization  and recombination rates make $\rho^{445}$ power-dependent. The variation of the equilibrium charge state with $P^{445}$ is a characteristic of wavelength regime B, which to the best of our knowledge has not yet been experimentally explored.
 
We also find evidence for a minor  contribution of two-photon ionization with 445 nm excitation. At high laser powers, $\rho(t_p)$ evolves in a bi-exponential manner featuring fast initial decay and slower recovery toward the steady-state value. Our three-level model predicts a bi-exponential evolution of $\rho(t_p)$ only if ionization is spin-dependent, \textit{i.e.} $k_{\mathrm{i,0}}\neq k_{\mathrm{i,1}}$. Notably, this spin-dependent two-photon ionization is possible due to non-zero PLE of NV$^{-}$ at this wavelength.\cite{Beha2012} The reduced, yet finite PLE of NV$^{-}$ under blue light illumination manifests itself in a  surprisingly high steady-state ODMR contrast of approximately 50\% of the original value, independent of $P^{445}$. A promising consequence of this result is that conventional ODMR experiments are compatible with continuous-wave illumination at 445 nm.

\textbf{Figure \ref{revfig2}c,d} presents an analogous investigation for wavelength regime A, using \SI{375}{nm} laser excitation (protocol IA). These data are recorded using a second NV. In contrast to our earlier findings for regime B, a mono-exponential decay of $\rho(t_p)$ at all laser powers to a power-independent steady state value of $\sim$20\% is evident for regime A. Again, these observations are well-explained by our three-state model. In regime A, recombination can now occur \textit{via} one-photon processes, making $\rho^{375}$ independent of laser power. Ionization is also spin-independent ($k_{i,0} = k_{i, 1}$) since it occurs solely \textit{via} one-photon processes. Vanishing $c$ from the remaining NV$^-$ population at all powers is indicative of a complete loss of spin polarization ($k_s$ = 0). Since ionization events do not necessarily preserve spin state, rapid one-photon charge-state interconversion driven by \SI{375}{nm} excitation destroys spin-state fidelity.

\begin{figure}
\includegraphics[width=1\textwidth]{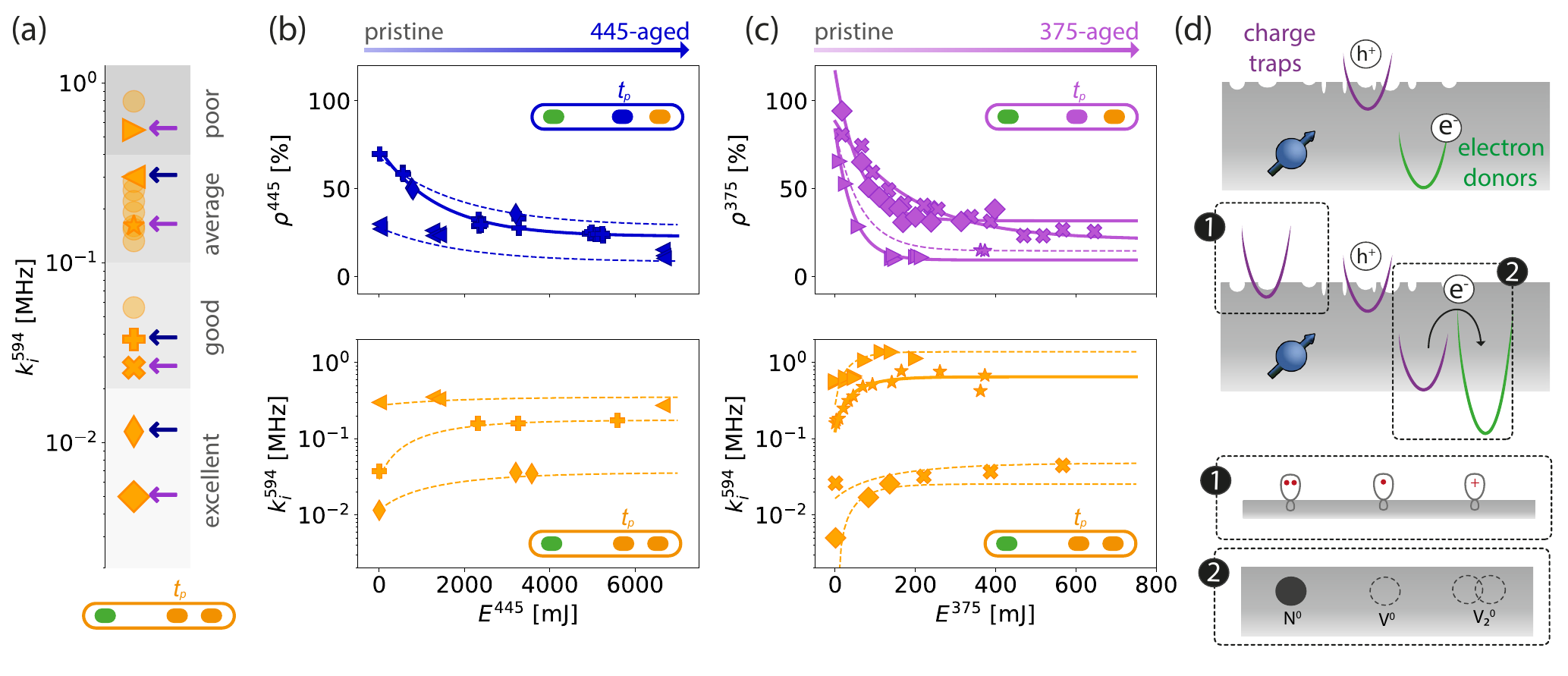}
\caption{\textbf{Aging of NV charge environments under prolonged exposure to 375 nm or 445 nm laser illumination.} \textbf{(a)} Ionization rates $k_i^{594}$ under orange laser illumination are utilized as classifiers for the local charge-environment of the NV (ranging from poor to excellent). Blue (violet) arrows indicate which NVs were selected to probe the effects of \SI{445}{nm} (\SI{375}{nm}) light exposure.   \textbf{(b, c)} In-depth analysis of the charge environment evolution induced by blue (\textit{b}) or UV (\textit{c}) exposure. Steady state NV$^-$ populations  $\rho^{445}$ ($\rho^{375}$)  and  ionization rates   $k_i^{594}$ evolve in an exponential manner as a function of exposure dose $E^{445}$ ($E^{375}$).  Solid lines indicate least-squares exponential fits of the experimental data. Dashed lines visualize the evolution in all cases where no direct fit could be performed.  \textbf{(d)} \textit{Top}: Additional formation of electron traps may occur via 1) modification of the diamond surface, 2) depletion of existing electron traps. \textit{Bottom}: Modification of the diamond surface could include formation of reactive carbanions, carbon radicals or carbocations. Substitutional nitrogen atoms\cite{Manson2005}, vacancies\cite{Luo22} and divacancies\cite{Miyamoto2023} are expected to be efficiently excited with UV-blue wavelengths. }
\label{revfig3}
\end{figure}

Notably, we observed that the photophysical parameters changed during prolonged UV or blue laser exposure. For data presented in \textbf{Figure \ref{revfig2}} and SI, defects had been exposed to sufficient doses of blue or UV laser light to reach stable photophysical behavior. These data deviate from results obtained on pristine NVs at the beginning of each measurement series. Prior to prolonged UV or blue exposure, NVs consistently exhibit higher $\rho$ and lower $k_i$ than after illumination (shown in black in \textbf{Figure \ref{revfig2}}). Rapid deterioration is especially pronounced for UV excitation, where initial characterization of NVs reveals $\rho$ $\geq$ 75\% following UV laser pulses (\textbf{Figure \ref{revfig2}c}), similar to earlier findings of comparably minor ionization under UV illumination.\cite{Chen2013} 

To understand this apparent "aging" of NVs and their electronic environment, we measure the ionization rates $k_i^{594}$ under orange laser illumination (protocol IC) as a function of the total amount of energy $E$ delivered by blue or UV laser excitation. This ionization rate is sensitive to the local charge environment of the emitter, which varies significantly for shallow NVs even with comparable depths.\cite{Dhomkar2018,Bluvstein2019, Yuan2020} Our observed values of $k_i^{594}$   span two orders of magnitude for the entire set of pristine NVs (\textbf{Figure \ref{revfig3}a}). For analysis, we group NVs according to their orange ionization rates in categories ranging from excellent to poor (low to high $k_i^{594}$) and select defects from each group for full characterization. We measure exponential increases in $k_i^{594}$ by up to one order-of-magnitude with increasing \textit{E} (\textbf{Figure \ref{revfig3}b,c}). Notably, $10\times$ greater \textit{E} is required to reach the aged state  under \SI{445}{nm} irradiation compared to \SI{375}{nm} irradiation.  As expected for both wavelengths, the increase in $k_i^{594}$ is accompanied by a decrease in the equilibrium NV$^-$ fraction $\rho^{\lambda}$ on the same energy scale. 

The acceleration of the ionization dynamics is indicative of the formation of novel electron trap states under prolonged blue and UV exposure. Two mechanisms could explain this behavior: 1) the generation of persistent reactive species (\textit{i.e.}, carbon radicals, carbanions, or carbocations) at the diamond surface\cite{Rodgers2024} that can modify charge stability \textit{via} band bending at the interface\cite{Neethirajan2023} or 2) a more rapid depletion of existing electron donors in favor of deeper electron traps that were previously inaccessible (\textbf{Figure \ref{revfig3}d}).  The former mechanism is relevant in the context of photochemical surface functionalization techniques driven by UV and blue illumination.\cite{Rodgers2024} Exploring whether the reduced NV stability persists even after the reactive species had undergone further reaction would provide insight into their contribution to NV aging. The latter mechanism is hypothesized based on efficient excitation of substitutional nitrogen donors (N$^0$)\cite{Manson2005}, vacancies (V$^0$)\cite{Luo22} and di-vacancies (V$_2^0$) \cite{Miyamoto2023} with short wavelengths of light, thereby redistributing holes and altering their accessibility. In combination, these mechanisms could contribute to destabilization including sudden drops of ODMR contrast, blinking, and even complete and irreversible loss of photoluminescence. Such effects are known to occur for shallow emitters,\cite{Sangtawesin2019, Bluvstein2019} yet the chemical or physical nature of responsible trap states remains poorly understood. Further studies are required to establish connections between UV- or blue-light-induced aging and the deterioration of NV properties seen during conventional sensing with green laser excitation. If the same changes in the NVs charge environments are responsible for laser-induced aging, our insights could provide guidance for stabilization of shallow defects.

\begin{figure}
\includegraphics[width=1\textwidth]{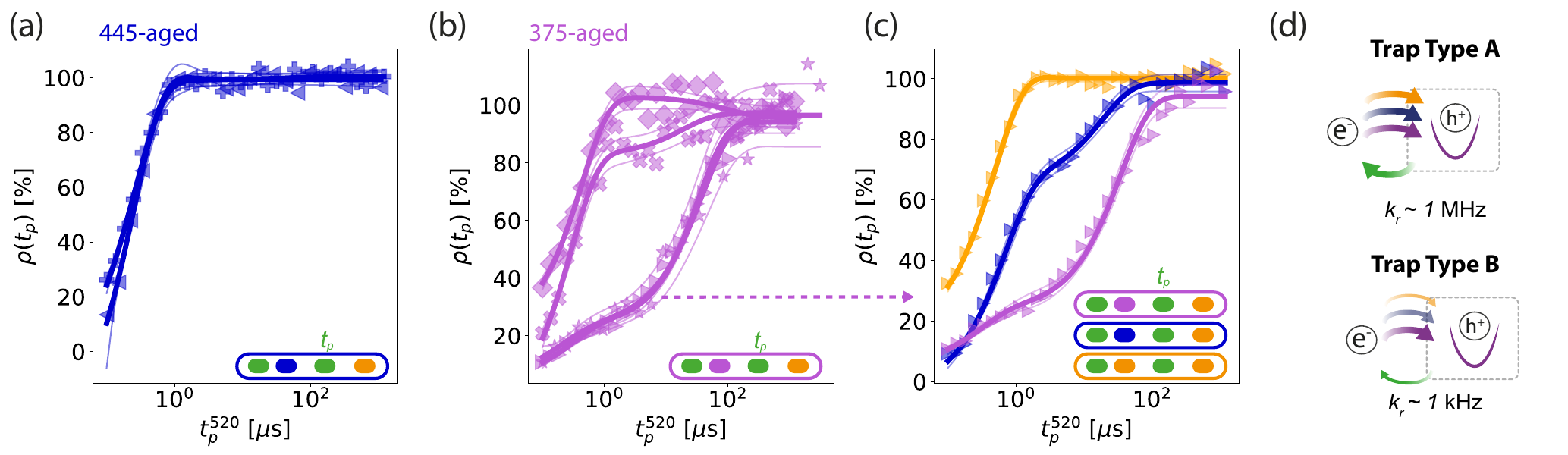}
\caption{\textbf{Re-initialization of the charge state using green illumination for aged NVs.} \textbf{(a)} Recombination after \SI{445}{nm} light induced ionization, measured on NVs that were aged with \SI{445}{nm} illumination. \textbf{(b)}  Recombination after \SI{375}{nm} light induced ionization, measured on NVs that were aged with \SI{375}{nm} illumination. To varying degree, the re-initialization now occurs at a ms timescale. \textbf{(c)}: For an NV showing slow recombination after UV induced ionization, the recombination measurement is repeated but using blue or orange light for ionization. \textbf{(d)} We propose two distinct electron trap states with different recombination rates under green illumination. Trap type A is present on pristine and aged NV centers, accessible at orange, blue and UV wavelengths and allows for fast recombination  under green illumination at a $\mu$s-timescale. Trap type B is only formed during UV-induced aging, only populated under UV or blue wavelengths and accounts for a slow recombination channel at a ms-timescale.}
\label{revfig4}
\end{figure}

To further analyze the nature of newly formed trap states, we study the charge recombination dynamics under green illumination (protocols IIA--C). Initial characterization of defects (\textit{i.e.}, before aging) reveals that recovery of the charge and spin states by green pulses is mono-exponential and exhibits rates $k_r^{520}$ $\approx$ \SI{1}{MHz}, indifferent to preceding perturbation by \SI{375}{nm}, \SI{445}{nm}, or \SI{594}{nm} pulses (SI Figure S4, S5, S6). After blue light-induced aging, these charge recombination dynamics remain unchanged as evidenced by mono-exponential recovery of $\rho$ on multiple NVs following ionization with 445 nm pulses (\textbf{Figure \ref{revfig4}a}). In contrast, UV-induced aging leads to modified charge recombination rates depending on the NV classification (excellent $\rightarrow$ poor). Recovery from UV-induced ionization becomes bi-exponential and slower for poor and, to a lesser extent, average, NVs (\textbf{Figure \ref{revfig3}b}). For these NVs, a steady-state $\rho(t_p)$ is only reached after $\approx$\SI{1}{ms}, corresponding to $k_r^{\prime} \approx$ \SI{1}{kHz}. The bi-exponential dynamics suggest the existence of a second recombination channel, consistent with the formation of a previously inaccessible electron trap. Importantly, these recombination rates depend on the wavelength of the preceding ionizing pulse. The fraction of the new, slowly recovering component is strongest for a \SI{375}{nm} pulse, still present for a \SI{445}{nm} pulse, and fully absent for ionization by \SI{594}{nm} pulses (\textbf{Figure \ref{revfig4}b}).  Thus, population of the novel electron traps following UV aging requires higher photon energies compared to the electron trap states existing on pristine NVs (\textbf{Figure \ref{revfig4}c}). Since recombination on blue-aged NVs following \SI{445}{nm} pulses do not feature this bi-exponential recovery behavior, we conclude that the types of traps formed upon prolonged blue \textit{versus} UV exposure are different.

\begin{figure}
\includegraphics[width=1.0\textwidth]{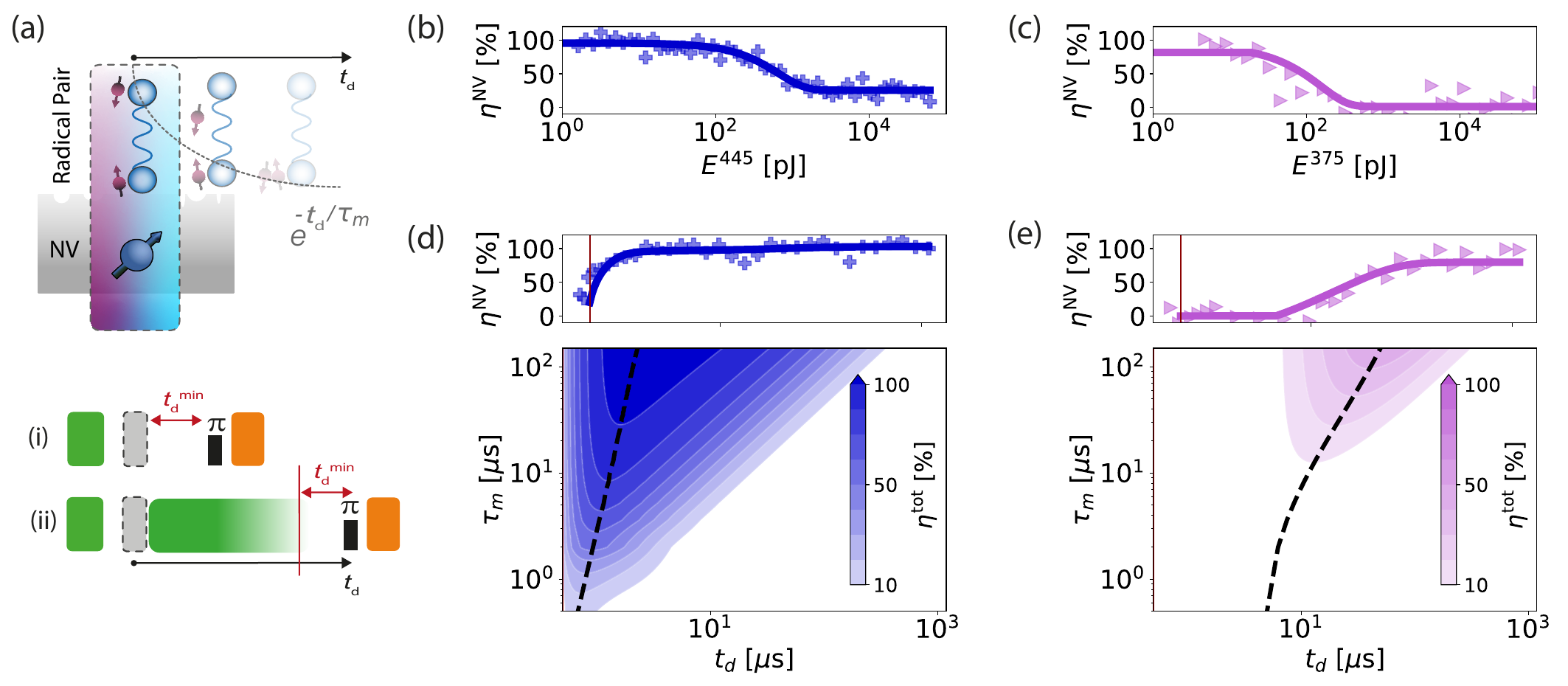}
\caption{\textbf{Consequences of blue and UV photo-induced spin and charge dynamics on sensing of photo-excitable molecular analytes.} \textbf{(a)} Schematic of the proposed experiment. The surface of a diamond sample containing shallow NVs is functionalized with molecules that support generation of radical pairs using UV or blue light at $t_d = 0$ following initialization of the NV with a green pulse. Population of the photo-generated radical pair species decays exponentially \textit{via} charge recombination with lifetime $\tau_m$. Sensing may be performed using two schemes \textit{i} and \textit{ii}; in the latter, the NV's charge and spin state are subsequently re-prepared using a second green laser pulse. In both, a microwave $\pi$ pulse and orange laser readout complete the ODMR sensing sequence. \textbf{(b,c)} Experimentally measured evolution of NV sensitivity $\eta^{NV}$ as a function of energy deposited by a blue (\textbf{b}) or UV (\textbf{c}) laser pulse on representative aged NVs.   
\textbf{(d,e)} \textit{Top:}  Experimentally measured recovery of sensitivity $\eta^{\mathrm{NV}}$ on the same NVs after a perturbing low-power \SI{445}{nm} (\SI{16}{\mu W} and  \SI{500}{ns}) (\textbf{b}) or \SI{375}nm (\SI{34}{\mu W}  and \SI{250}{\mu s}) (\textbf{c}) laser pulse. \textit{Bottom:} Total sensitivity $\eta^{\mathrm{tot}}$ calculated for radical pair lifetimes $\tau_m$ spanning \SI{500}{ns} to \SI{100}{\us} for blue (\textbf{d}) and UV (\textbf{e}) laser pulses. Black dashed lines follow the maximum $\eta^{\mathrm{tot}}$ for each scheme. }
\label{fig5} 
\end{figure} 

We finally assess the potential of near-surface NVs for probing photo-induced effects in proximal molecules using 445 nm or 375 nm excitation.\cite{Liu2017,Finkler2021,Voelker2023,Khurana2024} In the proposed experiment, molecules that support formation of photogenerated RPs, \textit{e.g.}, synthetic electron donor--bridge--acceptor species\cite{Chen2013} or biologically relevant blue-light sensitive cryptochromes,\cite{Biskup2009} are functionalized on the diamond surface close to a near-surface NV (\textbf{Figure \ref{fig5}a}).\cite{Abendroth2022,LiuBucher2022,Xie2022,Rodgers2024} After initialization of the NV, photoexcitation with a blue or UV pulse generates RPs in the molecules with lifetime $\tau_m$ before charge recombination. To measure the effect of these transient noise sources on the NV's ODMR contrast or frequency, a single microwave $\pi$-pulse at the NV $m_S = 0$ to $m_S = \pm1$ transition frequency follows the blue or UV pulses after delay $t_d$ with low-power orange readout pulse. 

A general expression for sensitivity of the NV in this experiment is given by $\eta^{\mathrm{NV}} \propto \sqrt{\rho} c$. \textbf{Figure \ref{fig5}b,c} shows experimentally measured evolution of $\eta^{\mathrm{NV}}$ as a function of the energy $E^{\lambda} = P^{\lambda}t_p$ delivered by a \SI{445}{nm} or \SI{375}{nm} laser pulse of power $P^{\lambda}$ and duration $t_p$ on two representative aged NVs. Promisingly, maximum $\eta^{\mathrm{NV}}$ is sustained for excitation pulses of both wavelengths that deliver less than \textit{ca.} \SI{10} {pJ}. Thus, for blue and UV pulses of sufficiently short duration or low power, normal ODMR measurements can be performed without the need for re-initialization and the minimum delay $t_d^{\mathrm{min}}$ is limited only by the NV$^-$ shelving state lifetime of \textit{ca.} 300 ns. Longer delays are required if the perturbing blue or UV pulse exceeding \SI{10}{pJ} scrambles NV charge and spin state or populates newly formed traps, necessitating re-initialization with a second green laser pulse. \textbf{Figure \ref{fig5}d,e} shows the total sensitivity in this regime, expressed as $\eta^{\mathrm{total}} \propto \eta^{\mathrm{NV}} e^{-t_d/\tau_m}$. In the case of 445 nm excitation, the non-zero PLE results in finite $\eta^{\mathrm{total}}$ even for $\tau_d=t_d^{\mathrm{min}}$. In comparison, after the 375 nm pulse, the slow recovery of $\eta^{\mathrm{NV}}$ would preclude sensing of RP species with $\tau_m \leq$ 10 $\mu$s.

In summary, we characterize the impact of 445 nm and 375 nm laser excitation on the charge and spin dynamics of near-surface NVs. The results agree well with the predicted UV and blue photophysics of NVs expected from a simplified three-level model. Our observations of power-dependent photoionization dynamics under blue illumination and accelerated deterioration of NV sensing properties under UV exposure provide important context to realize the potential of NVs to probe photochemical reactions and radical pair species. We identify a parameter space for both blue and UV laser excitation that would enable sensing of sufficiently long-lived photogenerated radical pairs in proximal molecules. Beyond photochemistry, the results herein provide insight into all-optical charge-state control of spin defects in diamond. Lastly, permanent aging of NVs proximal to surfaces is a serious challenge for long-term measurements with shallow defects, yet these effects have been described to date mostly anecdotally and remain poorly understood. This work takes an important step toward elucidating degradation of NV electronic environments by tracking the aging process.

\section{Supporting Information}
The Supporting Information provides details on materials and experimental methods, including additional data used to evaluate the conclusions in the paper; Figures S1--S5 and Tables S1 and S2 show the instrumentation schematics, charge and spin dynamics under blue and UV illumination for all NVs characterized, saturation curve of a representative NV under 445 nm illumination, and additional experimental details for data presented in the main text figures.

\section{Author Information}
\subsection{Corresponding Authors}

\textbf{Christian L. Degen}, \textit{Department of Physics, ETH Zürich, Otto-Stern-Weg 1, 8093 Zürich, Switzerland \& Quantum Center, ETH Zürich, 8093 Zürich, Switzerland}, ORCID: 0000-0003-2432-4301, E-mail: degenc@ethz.ch \newline
\textbf{John M. Abendroth}, \textit{Department of Physics, ETH Zürich, Otto-Stern-Weg 1, 8093 Zürich, Switzerland}, ORCID: 0000-0002-2369-4311, E-mail: jabendroth@phys.ethz.ch \subsection{Authors}
\textbf{Laura A. Völker}, \textit{Department of Physics, ETH Zürich, Otto-Stern-Weg 1, 8093 Zürich, Switzerland}, ORCID: 0000-0002-2990-9301 \newline
\textbf{Konstantin Herb}, \textit{Department of Physics, ETH Zürich, Otto-Stern-Weg 1, 8093 Zürich, Switzerland} %
\newline
\textbf{Darin A. Merchant}, \textit{Department of Physics, ETH Zürich, Otto-Stern-Weg 1, 8093 Zürich, Switzerland}  \newline
\textbf{Lorenzo Bechelli}, \textit{Department of Physics, ETH Zürich, Otto-Stern-Weg 1, 8093 Zürich, Switzerland}, ORCID: 0009-0003-4954-2679 \newline

\section{Author Contributions}
L.A.V., C.L.D., and J.M.A. conceived and designed the
experiments.
Data were collected by L.A.V., K.H., and D.A.M. All authors
discussed the results. The manuscript was written by L.A.V.,
C.L.D, and J.M.A. with assistance from all other authors.

\section{Notes}
The authors declare no competing financial interest.

\section{Acknowledgment}
The authors thank Stefan Ernst, Jan Rhensius, Erika Janitz and Tianqi Zhu for insightful discussions. J.M.A. acknowledges funding from the Swiss National Science Foundation (SNSF) Ambizione Project Grant No. PZ00P2-201590. C.L.D. acknowledges funding from the SBFI, Project "QMetMuFuSP" under Grant. No. UeM019-8, 215927 and from the SNSF, under Grants. No. 200021-219386 and CRSII-222812.

\bibliography{biblio}

\providecommand{\latin}[1]{#1}
\makeatletter
\providecommand{\doi}
  {\begingroup\let\do\@makeother\dospecials
  \catcode`\{=1 \catcode`\}=2 \doi@aux}
\providecommand{\doi@aux}[1]{\endgroup\texttt{#1}}
\makeatother
\providecommand*\mcitethebibliography{\thebibliography}
\csname @ifundefined\endcsname{endmcitethebibliography}
  {\let\endmcitethebibliography\endthebibliography}{}
\begin{mcitethebibliography}{49}
\providecommand*\natexlab[1]{#1}
\providecommand*\mciteSetBstSublistMode[1]{}
\providecommand*\mciteSetBstMaxWidthForm[2]{}
\providecommand*\mciteBstWouldAddEndPuncttrue
  {\def\EndOfBibitem{\unskip.}}
\providecommand*\mciteBstWouldAddEndPunctfalse
  {\let\EndOfBibitem\relax}
\providecommand*\mciteSetBstMidEndSepPunct[3]{}
\providecommand*\mciteSetBstSublistLabelBeginEnd[3]{}
\providecommand*\EndOfBibitem{}
\mciteSetBstSublistMode{f}
\mciteSetBstMaxWidthForm{subitem}{(\alph{mcitesubitemcount})}
\mciteSetBstSublistLabelBeginEnd
  {\mcitemaxwidthsubitemform\space}
  {\relax}
  {\relax}

\bibitem[Kothe \latin{et~al.}(1994)Kothe, Weber, Ohmes, Thurnauer, and
  Norris]{Kothe1994}
Kothe,~G.; Weber,~S.; Ohmes,~E.; Thurnauer,~M.~C.; Norris,~J.~R. High Time
  Resolution Electron Paramagnetic Resonance of Light-Induced Radical Pairs in
  Photosynthetic Bacterial Reaction Centers: Observation of Quantum Beats.
  \emph{J. Am. Chem. Soc.} \textbf{1994}, \emph{116}, 7729--7734\relax
\mciteBstWouldAddEndPuncttrue
\mciteSetBstMidEndSepPunct{\mcitedefaultmidpunct}
{\mcitedefaultendpunct}{\mcitedefaultseppunct}\relax
\EndOfBibitem
\bibitem[Levanon and Möbius(1997)Levanon, and Möbius]{Levanon1997}
Levanon,~H.; Möbius,~K. Advanced EPR Spectroscopy on Electron Transfer
  Processes in Photosynthesis and Biomimetic Model Systems. \emph{Annu. Rev.
  Biophys.} \textbf{1997}, \emph{26}, 495--540\relax
\mciteBstWouldAddEndPuncttrue
\mciteSetBstMidEndSepPunct{\mcitedefaultmidpunct}
{\mcitedefaultendpunct}{\mcitedefaultseppunct}\relax
\EndOfBibitem
\bibitem[Schmermund \latin{et~al.}(2019)Schmermund, Jurkaš, Özgen, Barone,
  Büchsenschütz, .~Winkler, Schmidt, Kourist, and Kroutil]{Schmermund2019}
Schmermund,~L.; Jurkaš,~V.; Özgen,~F.; Barone,~G.; Büchsenschütz,~H.;
  .~Winkler,~C.; Schmidt,~S.; Kourist,~R.; Kroutil,~W. Photo-Biocatalysis:
  Biotransformations in the Presence of Light. \emph{ACS Catal.} \textbf{2019},
  \emph{9}, 4115--4144\relax
\mciteBstWouldAddEndPuncttrue
\mciteSetBstMidEndSepPunct{\mcitedefaultmidpunct}
{\mcitedefaultendpunct}{\mcitedefaultseppunct}\relax
\EndOfBibitem
\bibitem[Hore and Mouritsen(2016)Hore, and Mouritsen]{Hore2016}
Hore,~P.; Mouritsen,~H. The Radical-Pair Mechanism of Magnetoreception.
  \emph{Annu. Rev. Biophys.} \textbf{2016}, \emph{45}, 299--344\relax
\mciteBstWouldAddEndPuncttrue
\mciteSetBstMidEndSepPunct{\mcitedefaultmidpunct}
{\mcitedefaultendpunct}{\mcitedefaultseppunct}\relax
\EndOfBibitem
\bibitem[Biskup \latin{et~al.}(2009)Biskup, Schleicher, Okafuji, Link, Hitomi,
  Getzoff, and Weber]{Biskup2009}
Biskup,~T.; Schleicher,~E.; Okafuji,~A.; Link,~G.; Hitomi,~K.; Getzoff,~E.;
  Weber,~S. Direct Observation of a Photoinduced Radical Pair in a Cryptochrome
  Blue-Light Photoreceptor. \emph{Angew. Chem. Int. Ed.} \textbf{2009},
  \emph{48}, 404--407\relax
\mciteBstWouldAddEndPuncttrue
\mciteSetBstMidEndSepPunct{\mcitedefaultmidpunct}
{\mcitedefaultendpunct}{\mcitedefaultseppunct}\relax
\EndOfBibitem
\bibitem[Hadi and Simon(2022)Hadi, and Simon]{Hadi2022}
Hadi,~Z.-H.; Simon,~C. Magnetic Field Effects in Biology from the Perspective
  of the Radical Pair Mechanism. \emph{J. R. Soc. Interface} \textbf{2022},
  \emph{19}, 20220325\relax
\mciteBstWouldAddEndPuncttrue
\mciteSetBstMidEndSepPunct{\mcitedefaultmidpunct}
{\mcitedefaultendpunct}{\mcitedefaultseppunct}\relax
\EndOfBibitem
\bibitem[Mani(2022)]{Mani2022}
Mani,~T. Molecular Qubits Based on Photogenerated Spin-Correlated Radical Pairs
  for Quantum Sensing. \emph{Chem. Phys. Rev.} \textbf{2022}, \emph{3},
  021301\relax
\mciteBstWouldAddEndPuncttrue
\mciteSetBstMidEndSepPunct{\mcitedefaultmidpunct}
{\mcitedefaultendpunct}{\mcitedefaultseppunct}\relax
\EndOfBibitem
\bibitem[Rugg \latin{et~al.}(2019)Rugg, Krzyaniak, Phelan, Ratner, Young, and
  Wasielewski]{Rugg2019}
Rugg,~B.; Krzyaniak,~M.; Phelan,~B.; Ratner,~M.; Young,~R.; Wasielewski,~M.
  Photodriven Quantum Teleportation of an Electron Spin State in a Covalent
  Donor-Acceptor-Radical System. \emph{Nat. Chem.} \textbf{2019}, \emph{11},
  981--986\relax
\mciteBstWouldAddEndPuncttrue
\mciteSetBstMidEndSepPunct{\mcitedefaultmidpunct}
{\mcitedefaultendpunct}{\mcitedefaultseppunct}\relax
\EndOfBibitem
\bibitem[Neleson \latin{et~al.}(2020)Neleson, Zhang, Zhou, Rugg, Krzyaniak, and
  Wasielewski]{Nelson2020}
Neleson,~J.; Zhang,~J.; Zhou,~J.; Rugg,~B.; Krzyaniak,~M.; Wasielewski,~M. CNOT
  Gate Operation on a Photogenerated Molecular Electron Spin-Qubit Pair.
  \emph{J. Chem. Phys.} \textbf{2020}, \emph{152}, 014503\relax
\mciteBstWouldAddEndPuncttrue
\mciteSetBstMidEndSepPunct{\mcitedefaultmidpunct}
{\mcitedefaultendpunct}{\mcitedefaultseppunct}\relax
\EndOfBibitem
\bibitem[Mao \latin{et~al.}(2023)Mao, Pažėra, Young, Krzyaniak, and
  Wasielewski]{Mao2023}
Mao,~H.; Pažėra,~G.~J.; Young,~R.~M.; Krzyaniak,~M.~D.; Wasielewski,~M.~R.
  Quantum Gate Operations on a Spectrally Addressable Photogenerated Molecular
  Electron Spin-Qubit Pair. \emph{J. Am. Chem. Soc.} \textbf{2023}, \emph{145},
  6585--6593\relax
\mciteBstWouldAddEndPuncttrue
\mciteSetBstMidEndSepPunct{\mcitedefaultmidpunct}
{\mcitedefaultendpunct}{\mcitedefaultseppunct}\relax
\EndOfBibitem
\bibitem[Liu \latin{et~al.}(2017)Liu, Plenio, and Cai]{Liu2017}
Liu,~H.; Plenio,~M.; Cai,~J. Scheme for Detection of Single-Molecule Radical
  Pair Reaction Using Spin in Diamond. \emph{Phys. Rev. Lett.} \textbf{2017},
  \emph{118}, 200402\relax
\mciteBstWouldAddEndPuncttrue
\mciteSetBstMidEndSepPunct{\mcitedefaultmidpunct}
{\mcitedefaultendpunct}{\mcitedefaultseppunct}\relax
\EndOfBibitem
\bibitem[Finkler and Dasari(2021)Finkler, and Dasari]{Finkler2021}
Finkler,~A.; Dasari,~D. Quantum Sensing and Control of Spin-State Dynamics in
  the Radical-Pair Mechanism. \emph{Phys. Rev. Appl.} \textbf{2021}, \emph{15},
  034066\relax
\mciteBstWouldAddEndPuncttrue
\mciteSetBstMidEndSepPunct{\mcitedefaultmidpunct}
{\mcitedefaultendpunct}{\mcitedefaultseppunct}\relax
\EndOfBibitem
\bibitem[Völker \latin{et~al.}(2023)Völker, Herb, Janitz, Degen, and
  Abendroth]{Voelker2023}
Völker,~L.~A.; Herb,~K.; Janitz,~E.; Degen,~C.~L.; Abendroth,~J.~M. Toward
  Quantum Sensing of Chiral Induced Spin Selectivity: Probing
  Donor-Bridge-Acceptor Molecules with NV Centers in Diamond. \emph{J. Chem.
  Phys.} \textbf{2023}, \emph{158}, 161103\relax
\mciteBstWouldAddEndPuncttrue
\mciteSetBstMidEndSepPunct{\mcitedefaultmidpunct}
{\mcitedefaultendpunct}{\mcitedefaultseppunct}\relax
\EndOfBibitem
\bibitem[Khurana \latin{et~al.}(2024)Khurana, Jensen, Giri, Bocquel, Andersen,
  Berg-S\o{}rensen, and Huck]{Khurana2024}
Khurana,~D.; Jensen,~R.~H.; Giri,~R.; Bocquel,~J.; Andersen,~U.~L.;
  Berg-S\o{}rensen,~K.; Huck,~A. Sensing of Magnetic Field Effects in
  Radical-Pair Reactions using a Quantum Sensor. \emph{Phys. Rev. Res.}
  \textbf{2024}, \emph{6}, 013218\relax
\mciteBstWouldAddEndPuncttrue
\mciteSetBstMidEndSepPunct{\mcitedefaultmidpunct}
{\mcitedefaultendpunct}{\mcitedefaultseppunct}\relax
\EndOfBibitem
\bibitem[Lovchinsky \latin{et~al.}(2016)Lovchinsky, Sushkov, Urbach, de~Leon,
  Choi, Greve, Evans, Gertner, Bersin, Müller, McGuinness, Jelezko, Walsworth,
  Park, and Lukin]{Lovchinsky_science_2016}
Lovchinsky,~I.; Sushkov,~A.~O.; Urbach,~E.; de~Leon,~N.~P.; Choi,~S.;
  Greve,~K.~D.; Evans,~R.; Gertner,~R.; Bersin,~E.; Müller,~C.;
  McGuinness,~L.; Jelezko,~F.; Walsworth,~R.~L.; Park,~H.; Lukin,~M.~D. Nuclear
  Magnetic Resonance Detection and Spectroscopy of Single Proteins using
  Quantum Logic. \emph{Science} \textbf{2016}, \emph{351}, 836--841\relax
\mciteBstWouldAddEndPuncttrue
\mciteSetBstMidEndSepPunct{\mcitedefaultmidpunct}
{\mcitedefaultendpunct}{\mcitedefaultseppunct}\relax
\EndOfBibitem
\bibitem[Janitz \latin{et~al.}(2022)Janitz, Herb, Völker, Huxter, Degen, and
  Abendroth]{Janitz2022}
Janitz,~E.; Herb,~K.; Völker,~L.~A.; Huxter,~W.~S.; Degen,~C.~L.;
  Abendroth,~J.~M. Diamond Surface Engineering for Molecular Sensing with
  Nitrogen—Vacancy Centers. \emph{J. Mater. Chem. C} \textbf{2022},
  \emph{10}, 13533--13569\relax
\mciteBstWouldAddEndPuncttrue
\mciteSetBstMidEndSepPunct{\mcitedefaultmidpunct}
{\mcitedefaultendpunct}{\mcitedefaultseppunct}\relax
\EndOfBibitem
\bibitem[Abendroth \latin{et~al.}(2022)Abendroth, Herb, Janitz, Zhu, Völker,
  and Degen]{Abendroth2022}
Abendroth,~J.~M.; Herb,~K.; Janitz,~E.; Zhu,~T.; Völker,~L.~A.; Degen,~C.~L.
  Single-Nitrogen-Vacancy NMR of Amine-Functionalized Diamond Surfaces.
  \emph{Nano Lett.} \textbf{2022}, \emph{22}, 7294--7303\relax
\mciteBstWouldAddEndPuncttrue
\mciteSetBstMidEndSepPunct{\mcitedefaultmidpunct}
{\mcitedefaultendpunct}{\mcitedefaultseppunct}\relax
\EndOfBibitem
\bibitem[Liu \latin{et~al.}(2022)Liu, Ma, Rizzato, Semrau, Henning, Sharp,
  Fischer, and Bucher]{LiuBucher2022}
Liu,~K.~S.; Ma,~X.; Rizzato,~R.; Semrau,~A.~L.; Henning,~A.; Sharp,~I.~D.;
  Fischer,~R.~A.; Bucher,~D.~B. Using Metal-Organic Frameworks to Confine
  Liquid Samples for Nanoscale NV-NMR. \emph{Nano Lett.} \textbf{2022},
  \emph{22}, 9876--9882\relax
\mciteBstWouldAddEndPuncttrue
\mciteSetBstMidEndSepPunct{\mcitedefaultmidpunct}
{\mcitedefaultendpunct}{\mcitedefaultseppunct}\relax
\EndOfBibitem
\bibitem[Xie \latin{et~al.}(2022)Xie, Yu, Rodgers, Xu, Chi-Durán, Toros,
  Quack, de~Leon, and Maurer]{Xie2022}
Xie,~M.; Yu,~X.; Rodgers,~L. V.~H.; Xu,~D.; Chi-Durán,~I.; Toros,~A.;
  Quack,~N.; de~Leon,~N.~P.; Maurer,~P.~C. Biocompatible Surface
  Functionalization Architecture for a Diamond Quantum Sensor. \emph{Proc.
  Natl. Acad. Sci. U. S. A.} \textbf{2022}, \emph{119}, e2114186119\relax
\mciteBstWouldAddEndPuncttrue
\mciteSetBstMidEndSepPunct{\mcitedefaultmidpunct}
{\mcitedefaultendpunct}{\mcitedefaultseppunct}\relax
\EndOfBibitem
\bibitem[Rodgers \latin{et~al.}(2024)Rodgers, Nguyen, Cox, Zervas, Yuan,
  Sangtawesin, Stacey, Jaye, Weiland, Pershin, Gali, Thomsen, Meynell, Hughes,
  Bleszynski~Jayich, Gui, Cava, Knowles, and de~Leon]{Rodgers2024}
Rodgers,~L. \latin{et~al.}  Diamond Surface Functionalization via Visible
  Light-Driven C-H Activation for Nanoscale Quantum Sensing. \emph{Proceedings
  of the National Academy of Sciences} \textbf{2024}, \emph{121},
  e2316032121\relax
\mciteBstWouldAddEndPuncttrue
\mciteSetBstMidEndSepPunct{\mcitedefaultmidpunct}
{\mcitedefaultendpunct}{\mcitedefaultseppunct}\relax
\EndOfBibitem
\bibitem[Perona~Martínez \latin{et~al.}(2020)Perona~Martínez, Nusantara,
  Chipaux, Padamati, and Schirhagl]{Martinez2020}
Perona~Martínez,~F.; Nusantara,~A.~C.; Chipaux,~M.; Padamati,~S.~K.;
  Schirhagl,~R. Nanodiamond Relaxometry-Based Detection of Free-Radical Species
  When Produced in Chemical Reactions in Biologically Relevant Conditions.
  \emph{ACS Sensors} \textbf{2020}, \emph{5}, 3862--3869\relax
\mciteBstWouldAddEndPuncttrue
\mciteSetBstMidEndSepPunct{\mcitedefaultmidpunct}
{\mcitedefaultendpunct}{\mcitedefaultseppunct}\relax
\EndOfBibitem
\bibitem[Ninio \latin{et~al.}(2021)Ninio, Waiskopf, Meirzada, Romach, Haim,
  Yochelis, Banin, and Bar-Gill]{Ninio2021}
Ninio,~Y.; Waiskopf,~N.; Meirzada,~I.; Romach,~Y.; Haim,~G.; Yochelis,~S.;
  Banin,~U.; Bar-Gill,~N. High-Sensitivity, High-Resolution Detection of
  Reactive Oxygen Species Concentration Using NV Centers. \emph{ACS Photonics}
  \textbf{2021}, \emph{8}, 1917--1921\relax
\mciteBstWouldAddEndPuncttrue
\mciteSetBstMidEndSepPunct{\mcitedefaultmidpunct}
{\mcitedefaultendpunct}{\mcitedefaultseppunct}\relax
\EndOfBibitem
\bibitem[Gaebel \latin{et~al.}(2006)Gaebel, Domhan, Wittmann, Popa, Jelezko,
  Rabeau, Greentree, Prawer, Trajkov, Hemmer, and Wrachtrup]{Gaebel2006}
Gaebel,~T.; Domhan,~M.; Wittmann,~C.; Popa,~I.; Jelezko,~F.; Rabeau,~J.;
  Greentree,~A.; Prawer,~S.; Trajkov,~E.; Hemmer,~P.~R.; Wrachtrup,~J.
  Photochromism in Single Nitrogen-Vacancy Defect in Diamond. \emph{Applied
  Physics B} \textbf{2006}, \emph{82}, 243--246\relax
\mciteBstWouldAddEndPuncttrue
\mciteSetBstMidEndSepPunct{\mcitedefaultmidpunct}
{\mcitedefaultendpunct}{\mcitedefaultseppunct}\relax
\EndOfBibitem
\bibitem[Waldherr \latin{et~al.}(2011)Waldherr, Beck, Steiner, Neumann, Gali,
  Frauenheim, Jelezko, and Wrachtrup]{Waldherr2011}
Waldherr,~G.; Beck,~J.; Steiner,~M.; Neumann,~P.; Gali,~A.; Frauenheim,~T.;
  Jelezko,~F.; Wrachtrup,~J. Dark States of Single Nitrogen-Vacancy Centers in
  Diamond Unraveled by Single Shot NMR. \emph{Phys. Rev. Lett.} \textbf{2011},
  \emph{106}, 157601\relax
\mciteBstWouldAddEndPuncttrue
\mciteSetBstMidEndSepPunct{\mcitedefaultmidpunct}
{\mcitedefaultendpunct}{\mcitedefaultseppunct}\relax
\EndOfBibitem
\bibitem[Beha \latin{et~al.}(2012)Beha, Batalov, Manson, Bratschitsch, and
  Leitenstorfer]{Beha2012}
Beha,~K.; Batalov,~A.; Manson,~N.~B.; Bratschitsch,~R.; Leitenstorfer,~A.
  Optimum Photoluminescence Excitation and Recharging Cycle of Single
  Nitrogen-Vacancy Centers in Ultrapure Diamond. \emph{Phys. Rev. Lett.}
  \textbf{2012}, \emph{109}, 097404\relax
\mciteBstWouldAddEndPuncttrue
\mciteSetBstMidEndSepPunct{\mcitedefaultmidpunct}
{\mcitedefaultendpunct}{\mcitedefaultseppunct}\relax
\EndOfBibitem
\bibitem[Aslam \latin{et~al.}(2013)Aslam, Waldherr, Neumann, Jelezko, and
  Wrachtrup]{Aslam2013}
Aslam,~N.; Waldherr,~G.; Neumann,~P.; Jelezko,~F.; Wrachtrup,~J. Photo-Induced
  Ionization Dynamics of the Nitrogen Vacancy Defect in Diamond Investigated by
  Single-Shot Charge State Detection. \emph{New J. Phys.} \textbf{2013},
  \emph{15}, 013604\relax
\mciteBstWouldAddEndPuncttrue
\mciteSetBstMidEndSepPunct{\mcitedefaultmidpunct}
{\mcitedefaultendpunct}{\mcitedefaultseppunct}\relax
\EndOfBibitem
\bibitem[Hacquebard and Childress(2018)Hacquebard, and
  Childress]{Hacquebard2018}
Hacquebard,~L.; Childress,~L. Charge-State Dynamics During Excitation and
  Depletion of the Nitrogen-Vacancy Center in Diamond. \emph{Phys. Rev. A}
  \textbf{2018}, \emph{97}, 063408\relax
\mciteBstWouldAddEndPuncttrue
\mciteSetBstMidEndSepPunct{\mcitedefaultmidpunct}
{\mcitedefaultendpunct}{\mcitedefaultseppunct}\relax
\EndOfBibitem
\bibitem[Razinkovas \latin{et~al.}(2021)Razinkovas, Maciaszek, Reinhard,
  Doherty, and Alkauskas]{Razinkovas2021}
Razinkovas,~L.; Maciaszek,~M.; Reinhard,~F.; Doherty,~M.~W.; Alkauskas,~A.
  Photoionization of Negatively Charged NV Centers in Diamond: Theory and ab
  initio Calculations. \emph{Phys. Rev. B} \textbf{2021}, \emph{104},
  235301\relax
\mciteBstWouldAddEndPuncttrue
\mciteSetBstMidEndSepPunct{\mcitedefaultmidpunct}
{\mcitedefaultendpunct}{\mcitedefaultseppunct}\relax
\EndOfBibitem
\bibitem[Han \latin{et~al.}(2010)Han, Kim, Eggeling, and Hell]{Han2010}
Han,~K.~Y.; Kim,~S.~K.; Eggeling,~C.; Hell,~S.~W. Metastable Dark States Enable
  Ground State Depletion Microscopy of Nitrogen Vacancy Centers in Diamond with
  Diffraction-Unlimited Resolution. \emph{Nano Lett.} \textbf{2010}, \emph{10},
  3199--3203\relax
\mciteBstWouldAddEndPuncttrue
\mciteSetBstMidEndSepPunct{\mcitedefaultmidpunct}
{\mcitedefaultendpunct}{\mcitedefaultseppunct}\relax
\EndOfBibitem
\bibitem[Chen \latin{et~al.}(2013)Chen, Zou, Sun, and Guo]{Chen2013}
Chen,~C.-D.; Zou,~C.-L.; Sun,~F.-W.; Guo,~G.-C. Optical Manipulation of the
  Charge State of Nitrogen-Vacancy Center in Diamond. \emph{Appl. Phys. Lett.}
  \textbf{2013}, \emph{103}, 013112\relax
\mciteBstWouldAddEndPuncttrue
\mciteSetBstMidEndSepPunct{\mcitedefaultmidpunct}
{\mcitedefaultendpunct}{\mcitedefaultseppunct}\relax
\EndOfBibitem
\bibitem[Bourgeois \latin{et~al.}(2017)Bourgeois, Londero, Buczak, Hruby,
  Gulka, Balasubramaniam, Wachter, Stursa, Dobes, Aumayr, Trupke, Gali, and
  Nesladek]{Bourgeois2017}
Bourgeois,~E.; Londero,~E.; Buczak,~K.; Hruby,~J.; Gulka,~M.;
  Balasubramaniam,~Y.; Wachter,~G.; Stursa,~J.; Dobes,~K.; Aumayr,~F.;
  Trupke,~M.; Gali,~A.; Nesladek,~M. Enhanced Photoelectric Detection of NV
  Magnetic Resonances in Diamond Under Dual-Beam Excitation. \emph{Phys. Rev.
  B} \textbf{2017}, \emph{95}, 041402\relax
\mciteBstWouldAddEndPuncttrue
\mciteSetBstMidEndSepPunct{\mcitedefaultmidpunct}
{\mcitedefaultendpunct}{\mcitedefaultseppunct}\relax
\EndOfBibitem
\bibitem[Li \latin{et~al.}(2022)Li, Zhang, Zhou, Hu, Ma, Zhang, and Yi]{Li2022}
Li,~C.; Zhang,~Q.; Zhou,~N.; Hu,~B.; Ma,~C.; Zhang,~C.; Yi,~Z. UV-Induced
  Charge-State Conversion from the Negatively to Neutrally Charged
  Nitrogen-Vacancy Centers in Diamond. \emph{J. Appl. Phys.} \textbf{2022},
  \emph{132}, 215102\relax
\mciteBstWouldAddEndPuncttrue
\mciteSetBstMidEndSepPunct{\mcitedefaultmidpunct}
{\mcitedefaultendpunct}{\mcitedefaultseppunct}\relax
\EndOfBibitem
\bibitem[Yang \latin{et~al.}(2022)Yang, Huang, Bista, Ding, Chen, and
  Chiang]{Yang2022}
Yang,~T.; Huang,~Y.-W.; Bista,~P.; Ding,~C.-F.; Chen,~J.;
  Chiang,~H.-C.,~C.-T.~Chang Photoluminescence of Nitrogen-Vacancy Centers by
  Ultraviolet One- and Two-Photon Excitation of Fluorescent Nanodiamonds.
  \emph{J. Phys. Chem. Lett.} \textbf{2022}, \emph{13}, 11280--11287\relax
\mciteBstWouldAddEndPuncttrue
\mciteSetBstMidEndSepPunct{\mcitedefaultmidpunct}
{\mcitedefaultendpunct}{\mcitedefaultseppunct}\relax
\EndOfBibitem
\bibitem[Wood \latin{et~al.}(2024)Wood, Lozovoi, Goldblatt, Meriles, and
  Martin]{Wood2024}
Wood,~A.~A.; Lozovoi,~A.; Goldblatt,~R.~M.; Meriles,~C.~A.; Martin,~A.~M.
  Wavelength Dependence of Nitrogen Vacancy Center Charge Cycling. \emph{Phys.
  Rev. B} \textbf{2024}, \emph{109}, 134106\relax
\mciteBstWouldAddEndPuncttrue
\mciteSetBstMidEndSepPunct{\mcitedefaultmidpunct}
{\mcitedefaultendpunct}{\mcitedefaultseppunct}\relax
\EndOfBibitem
\bibitem[Bradac \latin{et~al.}(2010)Bradac, Gaebel, Naidoo, Sellars, Twamley,
  Brown, Barnard, Plakhotnik, Zvyagin, and Rabeau]{Bradac2010}
Bradac,~C.; Gaebel,~T.; Naidoo,~N.; Sellars,~M.~J.; Twamley,~J.; Brown,~L.~J.;
  Barnard,~A.~S.; Plakhotnik,~T.; Zvyagin,~A.~V.; Rabeau,~J.~R. Observation and
  Control of Blinking Nitrogen-Vacancy Centres in Discrete Nanodiamonds.
  \emph{Nature Nanotechnology} \textbf{2010}, \emph{5}, 345--349\relax
\mciteBstWouldAddEndPuncttrue
\mciteSetBstMidEndSepPunct{\mcitedefaultmidpunct}
{\mcitedefaultendpunct}{\mcitedefaultseppunct}\relax
\EndOfBibitem
\bibitem[Dhomkar \latin{et~al.}(2018)Dhomkar, Jayakumar, Zangara, and
  Meriles]{Dhomkar2018}
Dhomkar,~S.; Jayakumar,~H.; Zangara,~P.; Meriles,~C. Charge Dynamics in
  Near-Surface, Variable-Density Ensembles of Nitrogen-Vacancy Centers in
  Diamond. \emph{Nano Lett.} \textbf{2018}, \emph{18}, 4046--4052\relax
\mciteBstWouldAddEndPuncttrue
\mciteSetBstMidEndSepPunct{\mcitedefaultmidpunct}
{\mcitedefaultendpunct}{\mcitedefaultseppunct}\relax
\EndOfBibitem
\bibitem[Sangtawesin \latin{et~al.}(2019)Sangtawesin, Dwyer, Srinivasan,
  Allred, Rodgers, De~Greve, Stacey, Dontschuk, O'Donnell, Hu, Evans, Jaye,
  Fischer, Markham, Twitchen, Park, Lukin, and de~Leon]{Sangtawesin2019}
Sangtawesin,~S. \latin{et~al.}  Origins of Diamond Surface Noise Probed by
  Correlating Single-Spin Measurements with Surface Spectroscopy. \emph{Phys.
  Rev. X} \textbf{2019}, \emph{9}, 031052\relax
\mciteBstWouldAddEndPuncttrue
\mciteSetBstMidEndSepPunct{\mcitedefaultmidpunct}
{\mcitedefaultendpunct}{\mcitedefaultseppunct}\relax
\EndOfBibitem
\bibitem[Bluvstein \latin{et~al.}(2019)Bluvstein, Zhang, and
  Jayich]{Bluvstein2019}
Bluvstein,~D.; Zhang,~Z.; Jayich,~A. C.~B. Identifying and Mitigating Charge
  Instabilities in Shallow Diamond Nitrogen-Vacancy Centers. \emph{Phys. Rev.
  Lett.} \textbf{2019}, \emph{122}, 076101\relax
\mciteBstWouldAddEndPuncttrue
\mciteSetBstMidEndSepPunct{\mcitedefaultmidpunct}
{\mcitedefaultendpunct}{\mcitedefaultseppunct}\relax
\EndOfBibitem
\bibitem[Yuan \latin{et~al.}(2020)Yuan, Fitzpatrick, Rodgers, Sangtawesin,
  Srinivasan, and de~Leon]{Yuan2020}
Yuan,~Z.; Fitzpatrick,~M.; Rodgers,~L. V.~H.; Sangtawesin,~S.; Srinivasan,~S.;
  de~Leon,~N.~P. Charge State Dynamics and Optically Detected Electron Spin
  Resonance Contrast of Shallow Nitrogen-Vacancy Centers in Diamond.
  \emph{Phys. Rev. Res.} \textbf{2020}, \emph{2}, 033263\relax
\mciteBstWouldAddEndPuncttrue
\mciteSetBstMidEndSepPunct{\mcitedefaultmidpunct}
{\mcitedefaultendpunct}{\mcitedefaultseppunct}\relax
\EndOfBibitem
\bibitem[Shields \latin{et~al.}(2015)Shields, Unterreithmeier, de~Leon, Park,
  and Lukin]{Shields2015}
Shields,~B.~J.; Unterreithmeier,~Q.~P.; de~Leon,~N.~P.; Park,~H.; Lukin,~M.~D.
  Efficient Readout of a Single Spin State in Diamond via Spin-to-Charge
  Conversion. \emph{Phys. Rev. Lett.} \textbf{2015}, \emph{114}, 136402\relax
\mciteBstWouldAddEndPuncttrue
\mciteSetBstMidEndSepPunct{\mcitedefaultmidpunct}
{\mcitedefaultendpunct}{\mcitedefaultseppunct}\relax
\EndOfBibitem
\bibitem[Hopper \latin{et~al.}(2018)Hopper, Grote, Parks, and
  Bassett]{Hopper2018}
Hopper,~D.; Grote,~R.; Parks,~S.; Bassett,~L. Amplified Sensitivity of
  Nitrogen-Vacancy Spins in Nanodiamonds Using All-Optical Charge Readout.
  \emph{ACS Nano} \textbf{2018}, \emph{12}, 4678--4686\relax
\mciteBstWouldAddEndPuncttrue
\mciteSetBstMidEndSepPunct{\mcitedefaultmidpunct}
{\mcitedefaultendpunct}{\mcitedefaultseppunct}\relax
\EndOfBibitem
\bibitem[Ernst \latin{et~al.}(2023)Ernst, Scheidegger, Diesch, Lorenzelli, and
  Degen]{Ernst2023}
Ernst,~S.; Scheidegger,~P.~J.; Diesch,~S.; Lorenzelli,~L.; Degen,~C.~L.
  Temperature Dependence of Photoluminescence Intensity and Spin Contrast in
  Nitrogen-Vacancy Centers. \emph{Phys. Rev. Lett.} \textbf{2023}, \emph{131},
  086903\relax
\mciteBstWouldAddEndPuncttrue
\mciteSetBstMidEndSepPunct{\mcitedefaultmidpunct}
{\mcitedefaultendpunct}{\mcitedefaultseppunct}\relax
\EndOfBibitem
\bibitem[De\'ak \latin{et~al.}(2014)De\'ak, Aradi, Kaviani, Frauenheim, and
  Gali]{Deak2014}
De\'ak,~P.; Aradi,~B.; Kaviani,~M.; Frauenheim,~T.; Gali,~A. Formation of NV
  Centers in Diamond: A Theoretical Study Based on Calculated Transitions and
  Migration of Nitrogen and Vacancy Related Defects. \emph{Phys. Rev. B}
  \textbf{2014}, \emph{89}, 075203\relax
\mciteBstWouldAddEndPuncttrue
\mciteSetBstMidEndSepPunct{\mcitedefaultmidpunct}
{\mcitedefaultendpunct}{\mcitedefaultseppunct}\relax
\EndOfBibitem
\bibitem[Zhu \latin{et~al.}(2023)Zhu, Rhensius, Herb, Damle, Puebla-Hellmann,
  Degen, and Janitz]{Zhu2023}
Zhu,~T.; Rhensius,~J.; Herb,~K.; Damle,~V.; Puebla-Hellmann,~G.; Degen,~C.~L.;
  Janitz,~E. Multicone Diamond Waveguides for Nanoscale Quantum Sensing.
  \emph{Nano Lett.} \textbf{2023}, \emph{23}, 10110--10117\relax
\mciteBstWouldAddEndPuncttrue
\mciteSetBstMidEndSepPunct{\mcitedefaultmidpunct}
{\mcitedefaultendpunct}{\mcitedefaultseppunct}\relax
\EndOfBibitem
\bibitem[Manson and Harrison(2005)Manson, and Harrison]{Manson2005}
Manson,~N.; Harrison,~J. Photo-Ionization of the Nitrogen-Vacancy Center in
  Diamond. \emph{Diam. Relat. Mater.} \textbf{2005}, \emph{14},
  1705--1710\relax
\mciteBstWouldAddEndPuncttrue
\mciteSetBstMidEndSepPunct{\mcitedefaultmidpunct}
{\mcitedefaultendpunct}{\mcitedefaultseppunct}\relax
\EndOfBibitem
\bibitem[Luo \latin{et~al.}(2022)Luo, Lindner, Langer, Cimalla, Vidal, Hahl,
  Schreyvogel, Onoda, Ishii, and Ohshima]{Luo22}
Luo,~T.; Lindner,~L.; Langer,~J.; Cimalla,~V.; Vidal,~X.; Hahl,~F.;
  Schreyvogel,~C.; Onoda,~S.; Ishii,~S.; Ohshima,~T. Creation of
  Nitrogen-Vacancy Centers in Chemical Vapor Deposition Diamond for Sensing
  Applications. \emph{New J. Phys.} \textbf{2022}, \emph{24}, 033030\relax
\mciteBstWouldAddEndPuncttrue
\mciteSetBstMidEndSepPunct{\mcitedefaultmidpunct}
{\mcitedefaultendpunct}{\mcitedefaultseppunct}\relax
\EndOfBibitem
\bibitem[Miyamoto(2023)]{Miyamoto2023}
Miyamoto,~Y. Decay Process of Photoexcited Divacancies in Diamond Studied by
  First-Principles Simulations. \emph{Phys. Rev. Mater.} \textbf{2023},
  \emph{7}, 086002\relax
\mciteBstWouldAddEndPuncttrue
\mciteSetBstMidEndSepPunct{\mcitedefaultmidpunct}
{\mcitedefaultendpunct}{\mcitedefaultseppunct}\relax
\EndOfBibitem
\bibitem[Neethirajan \latin{et~al.}(2023)Neethirajan, Hache, Paone, Pinto,
  Denisenko, Stöhr, Udvarhelyi, Pershin, Gali, Wrachtrup, Kern, and
  Singha]{Neethirajan2023}
Neethirajan,~J.; Hache,~T.; Paone,~D.; Pinto,~D.; Denisenko,~A.; Stöhr,~R.;
  Udvarhelyi,~P.; Pershin,~A.; Gali,~A.; Wrachtrup,~J.; Kern,~K.; Singha,~A.
  Controlled Surface Modification to Revive Shallow NV$^-$ Centers. \emph{Nano
  Lett.} \textbf{2023}, \emph{23}, 2563--2569\relax
\mciteBstWouldAddEndPuncttrue
\mciteSetBstMidEndSepPunct{\mcitedefaultmidpunct}
{\mcitedefaultendpunct}{\mcitedefaultseppunct}\relax
\EndOfBibitem
\end{mcitethebibliography}


\providecommand{\latin}[1]{#1}
\makeatletter
\providecommand{\doi}
  {\begingroup\let\do\@makeother\dospecials
  \catcode`\{=1 \catcode`\}=2 \doi@aux}
\providecommand{\doi@aux}[1]{\endgroup\texttt{#1}}
\makeatother
\providecommand*\mcitethebibliography{\thebibliography}
\csname @ifundefined\endcsname{endmcitethebibliography}
  {\let\endmcitethebibliography\endthebibliography}{}
\begin{mcitethebibliography}{3}
\providecommand*\natexlab[1]{#1}
\providecommand*\mciteSetBstSublistMode[1]{}
\providecommand*\mciteSetBstMaxWidthForm[2]{}
\providecommand*\mciteBstWouldAddEndPuncttrue
  {\def\EndOfBibitem{\unskip.}}
\providecommand*\mciteBstWouldAddEndPunctfalse
  {\let\EndOfBibitem\relax}
\providecommand*\mciteSetBstMidEndSepPunct[3]{}
\providecommand*\mciteSetBstSublistLabelBeginEnd[3]{}
\providecommand*\EndOfBibitem{}
\mciteSetBstSublistMode{f}
\mciteSetBstMaxWidthForm{subitem}{(\alph{mcitesubitemcount})}
\mciteSetBstSublistLabelBeginEnd
  {\mcitemaxwidthsubitemform\space}
  {\relax}
  {\relax}

\bibitem[Ziegler \latin{et~al.}(2010)Ziegler, Ziegler, and Biersack]{SRIM}
Ziegler,~J.; Ziegler,~M.; Biersack,~J. SRIM -- The Stopping and Range of Ions
  in Matter (2010). \emph{Nucl. Instrum. Methods Phys. Res. B: Beam Interact.
  Mater. At.} \textbf{2010}, \emph{268}, 1818--1823\relax
\mciteBstWouldAddEndPuncttrue
\mciteSetBstMidEndSepPunct{\mcitedefaultmidpunct}
{\mcitedefaultendpunct}{\mcitedefaultseppunct}\relax
\EndOfBibitem
\bibitem[Zhu \latin{et~al.}(2023)Zhu, Rhensius, Herb, Damle, Puebla-Hellmann,
  Degen, and Janitz]{Zhu2023}
Zhu,~T.; Rhensius,~J.; Herb,~K.; Damle,~V.; Puebla-Hellmann,~G.; Degen,~C.~L.;
  Janitz,~E. Multicone Diamond Waveguides for Nanoscale Quantum Sensing.
  \emph{Nano Lett.} \textbf{2023}, \emph{23}, 10110--10117\relax
\mciteBstWouldAddEndPuncttrue
\mciteSetBstMidEndSepPunct{\mcitedefaultmidpunct}
{\mcitedefaultendpunct}{\mcitedefaultseppunct}\relax
\EndOfBibitem
\end{mcitethebibliography}

\end{document}


\newpage
\section{Materials and Methods}

\subsection{Diamond Sample Preparation}

A commercially available diamond membrane with nanofabricated pillar arrays containing shallow NV centers was used in this study (QZabre Ltd, Zürich, CH). The membrane is a quantum-grade (001)-terminated single-crystalline CVD-grown diamond (Element Six, Didcot, UK). The membrane was implanted with $^{15}$N$^{+}$ ions at 7 keV energy (CuttingEdge Ions, LLC; Anaheim, CA, USA) yielding an expected mean implantation depth of 10.8(40) nm determined using The Stopping and Range of Ions in Matter (SRIM) simulations\cite{SRIM}. The membrane was annealed at 880 \textdegree C for 2 h to convert implanted nitrogen into NV centers due to vacancy migration. Following annealing, nanopillar waveguide arrays hosting nominally single NV centers were fabricated by electron-beam lithography and reactive-ion etching.\cite{Zhu2023} The as-received membrane was cleaned for \textit{ca}. 2 h in a 1:1:1 tri-acid mixture of H$_{2}$SO$_{4}$:HClO$_{4}$:HNO$_{3}$ at 120 \textdegree C, baked at 460 \textdegree C under oxygen atmosphere (AS-Micro, RTP-System, Annealsys; Montpellier, FR) for 4 h, and then cleaned in piranha solution (3:1 H$_{2}$SO$_{4}$:H$_{2}$O$_{2}$) before measuring.

\subsection{Confocal Microscope for NV Measurements and Microwave Instrumentation}
\begin{figure} [H]
\includegraphics[width=0.8\textwidth]{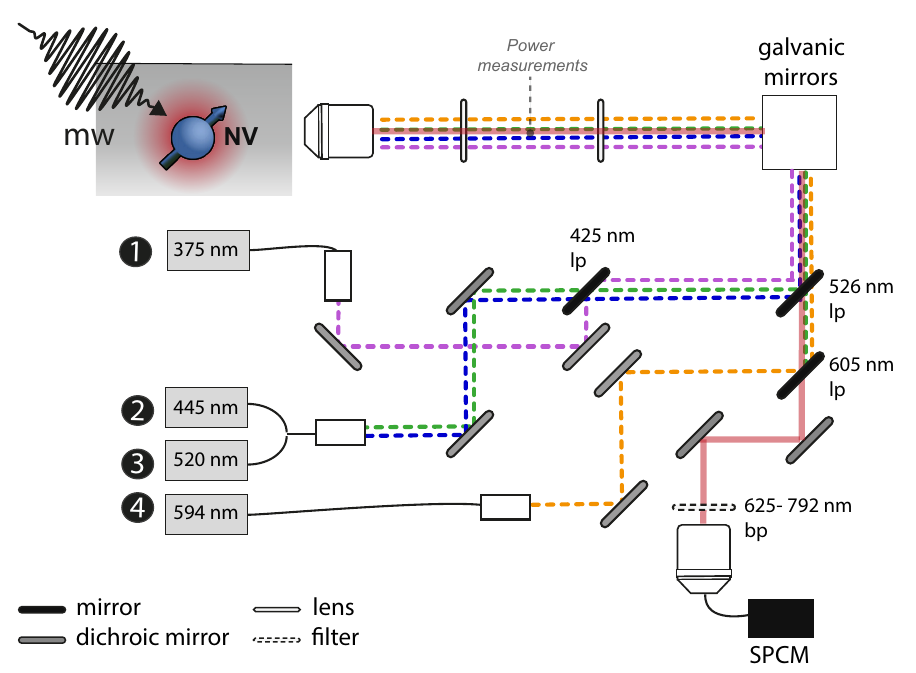}
\caption{Schematic layout of home-built confocal microscope utilized in this study. For dichroic mirrors and filters, relevant wavelengths are specified (lp = longpass, bp = bandpass). }
\label{setup}
\end{figure}

Experiments were performed on a custom-built confocal microscope, equipped with four different lasers at \SI{375}{nm} (Toptica iBeam Smart 375 diode laser), \SI{445}{nm} (homebuilt laser using diodes purchased from Frankfurt Laser), \SI{520}{nm}  (homebuilt laser using diode purchased from Frankfurt Laser) and \SI{594}{nm} (Cobolt Mambo 594 diode-pumped solid-state laser) and a 625--\SI{800}{nm} detection path. A schematic depiction of the setup is provided in Figure \ref{setup}. Optical pulses were generated either by using an acousto optic modulator (\SI{594}{nm}) or by direct modulation of diode drive current (\SI{375}{nm}, \SI{445}{nm}, \SI{520}{nm}). Lasers were focused on the sample with a high NA objective (Olympuys UPLXAPO 40$\times$). Photons emitted from the diamond sample were detected with a single-photon avalanche photodiode (Excelitas SPCM-AQRH family), time-tagged   (NI-PCIe-6363) and software-binned.  

Microwave pulses were synthesized using an arbitrary-waveform generator (Spectrum Instrumentation DN2.663-04) and up-converted by a vector signal generator (Stanford Research Systems SG-386). They were amplified (Mini Circuits High Power Amplifier ZHL-15W-422-S+) and delivered to the sample through a home-built coplanar waveguide. A static bias field $B_0$  was generated using a cylindrical samarium cobalt permanent magnet (TC-SmCo). Alignment of $B_0$ with the NV centers ZFS crystal axis was achieved by physically moving the magnet.

\subsection{Estimation of Laser Powers}
Laser powers reported in the main text refer to power at the sample. These values were estimated by measuring the power (Thorlabs C-series power sensor with S120VC sensor head) at the latest accessible point in the optical path (middle of 4f-system, see Figure \ref{setup}) and correcting for losses through the remaining components using the transmission data provided by the respective manufacturers. 

\subsection{NV Center Characterization}
NV centers were selected in nanopillar waveguide arrays hosting nominally single emitters  based on sufficiently high fluorescence intensity under green laser illumination. Resonance frequencies were determined by continuous-wave optically detected magnetic resonance (cw-ODMR) experiments and microwave pulse lengths were calibrated \textit{via} Rabi-measurements.  All measurements of this study were performed at a magnetic bias field of $\approx$ \SI{10}{mT} (\textit{i.e.}, resonance frequencies of $\approx$ \SI{2.59}{GHz}) and at Rabi-frequencies of $\approx$ \SI{7}{MHz}.   Laser pulses were followed by \SI{2}{\us} delays.   Charge and spin state init was performed using \SI{0.08}{mW} green laser illumination. Readout was performed using \SI{0.3}{mW} orange illumination and  fluorescence counts were integrated for \SI{300}{ns}. Ionization dynamics were characterized by  pulsed ODMR measurements at the NV centers resonance frequency for a series of ionizing pulses of varying length $t_p$ (see main text Figure 1(d) for pulse sequences).  Reference measurements without the ionizing pulse were taken simultaneously. Other relevant parameters are provided in Table \ref{measdetails}.

\small
\begin{longtable}[H]{p{1cm}|p{2.5cm}|p{10cm}} 
\caption{Experimental details (pulse length, laser powers, \textit{etc}.) for measurement data  shown in main text figures.  } \\
Fig. & Panel  & Details    \\  \hline\hline
2 & a, b &  Blue ionization as a function of blue laser pulse length: NV  was initialized with a \SI{15}{\us} green laser pulse. Ionization under \SI{445}{nm} illumination was probed for pulse lengths ranging from \SI{100}{ns} -- \SI{4}{ms} and laser powers  ranging from \SI{0.1}{mW} to $>$\SI{1}{mW}.  \\
 & c, d & UV ionization as a function of UV laser pulse length: NV was initialized with a \SI{250}{\us} green laser pulse. Ionization under \SI{375}{nm} illumination was probed for pulse lengths ranging from \SI{100}{ns} -- \SI{2}{ms} and laser powers  ranging from \SI{0.01}{mW} to $>$\SI{0.1}{mW}. \\ \hline
 3 & a, b bottom, c bottom &  Ionization rates under orange laser illumination: NV centers were initialized with \SI{15}{\us} green laser pulses and ionized using orange laser pulses of varying lengths, ranging from \SI{100}{ns} to several ms. Ionization rates were extracted via least squares fitting.   \\
 & b top &  Steady state NV$^-$ fraction remaining under blue laser illumination.  NVs  were initialized with a \SI{15}{\us} green laser pulse. Ionization was induced using a \SI{0.016}{mW} \SI{445}{nm} laser pulse of sufficient length (determined in ionization measurement, typically $\geq$ \SI{250}{\us}).  \\
 & c top &   Steady state NV$^-$ fraction remaining under UV laser illumination. NVs  were initialized with a $>$ \SI{250}{\us} green laser pulse (longer pulses required for aged NV centers due to altered recombination). Ionization was induced using a \SI{375}{nm}  (powers ranging from \SI{0.024}{mW} to \SI{0.069}{mW})  laser pulse of sufficient length (determined in ionization measurement, typically $\geq$ \SI{250}{\us}).  \\ \hline
 4 & a &  Recombination after blue induced illumination on NV centers that had been aged using blue laser light.  NVs were fully ionized using \SI{500}{\us}, \SI{0.016}{mW} \SI{445}{nm} laser pulses (Ionizing pulse length and power were chosen based on ionization measurement). Recombination was performed using green laser pulses with pulse lengths ranging from \SI{100}{ns} -- \SI{2}{ms}.  \\ 

 & b & Recombination after UV induced illumination on NV centers that had been aged using UV laser light. NV centers were fully ionized using \SI{250}{\us}, \SI{0.024}{mW} to \SI{0.069}{mW} \SI{375}{nm} illumination.   (Ionizing pulse length and power were chosen based on ionization measurement). Recombination was performed using green laser pulses with pulse length ranging from \SI{100}{ns} -- \SI{2}{ms}. \\
  & c & Recombination after i) orange, ii) blue and iii) UV induced illumination, measured on the same NV center  after UV induced aging. The NV center was ionized using 
 sufficiently long ($>$\SI{250}{us}) pulses of the respective color. In each case pulse length was chosen based on an ionization measurement. For  \SI{445}{nm}, ionization was performed at \SI{0.010}{mW}, for UV illumination at \SI{0.034}{mW}. The recombination was performed using green laser pulses with pulse lengths ranging from \SI{100}{ns} -- 
\SI{2}{ms}. \\ \hline
 5 & b &  NV sensitivity evolution as a function of blue exposure dose. Sensitivity was evaluated based on an ionization measurement performed under the following conditions. The NV  was initialized with a \SI{15}{\us} green laser pulse. Ionization under \SI{445}{nm} illumination was probed for pulse lengths ranging from \SI{100}{ns} -- \SI{4}{ms} at a laser powers of \SI{0.016}{mW}. \\

   & c &  NV sensitivity evolution as a function of UV exposure dose. Sensitivity was evaluated based on an ionization measurement performed under the following conditions. The NV  was initialized with a \SI{1000}{\us} green laser pulse. Ionization under \SI{375}{nm} illumination was probed for pulse lengths ranging from \SI{100}{ns} -- \SI{4}{ms} at a laser power of \SI{0.034}{mW}.\\
   & d &  NV sensitivity evolution as a function of green recombination pulse length. The sensitivity was evaluated based on a recombination measurement performed under the following conditions. The NV  was fully ionized with a \SI{500}{\us}, \SI{0.016}{mW}  blue laser pulse. The recombination was performed using green laser pulses with pulse lengths ranging from \SI{100}{ns} -- \SI{2}{ms}. \\
   & e &  NV sensitivity evolution as a function of green recombination pulse length. The sensitivity was evaluated based on a recombination measurement performed under the following conditions. The NV  was fully ionized with a \SI{250}{\us}, \SI{0.034}{mW} UV laser pulse. The recombination was performed using green laser pulses with pulse lengths ranging from \SI{100}{ns} -- \SI{2}{ms}.
 
\label{measdetails}
\end{longtable}

\subsection{Three Level Rate Model}
To physically motivate the mathematical form of the acquired ionization/recombination traces, we utilize a simplified three level model, consisting of two levels ($M_0$ and $M_1$) for spin states $m_S = 0$ and $m_S = \pm1$, respectively for NV$^{-}$ and a single level ($Z$) for  NV$^{0}$.  Charge and spin dynamics are parameterized using spin-dependent ionization rates  rates $k_{\mathrm{i},0}$ and $k_{\mathrm{i},1}$ from the $m_S = 0$ and $m_S = \pm 1$ spin states,  spin state polarization with a rate $k_{\mathrm{s}}$ and recombination at a  rate $k_{\mathrm{r}}$. Since ionization events do not necessarily preserve the spin state, we assume that reionization equally populates all spin levels of NV$^{-}$. The evolution of level populations is a system of three coupled differential equations, 
\begin{equation}
\frac{d \Vec{N}}{dt} = \begin{pmatrix}
-k_{\mathrm{i},0} & k_{\mathrm{s}} & k_{\mathrm{r}}\\
0 & -k_{\mathrm{i},1} - k_{\mathrm{s}}  & 2 k_{\mathrm{r}}\\
k_{\mathrm{i},0} & k_{\mathrm{i},1} & -3k_{\mathrm{r}}\\
\end{pmatrix} \cdot \begin{pmatrix}
M_0 \\
M_1 \\
Z \\
\end{pmatrix}.
\end{equation}

Starting from a negatively charged NV center with fully polarized spin state, \textit{i.e.}, $\Vec{N}(0) = (1, 0,0)$ the population of NV$^0$ evolves in a bi-exponential manner,
\begin{equation}
Z(t_p) = \alpha e^{-t/\tau_1}  + \beta e^{-t/\tau_2} + \gamma
\end{equation}
with 
\begin{equation}
\tau_1 = \frac{1}{(k_{i,0} + k_s + k_{i,1} + 3k_r) + k_w}
\end{equation}
\begin{equation}
\tau_2 = \frac{1}{(k_{i,0} + k_s + k_{i,1} + 3k_r) - k_w}
\end{equation}
where we have introduced  
\begin{equation}
k_w =   \sqrt{(-k_{i,0} + k_s + k_{i,1})^2 - 2(k_{i,0} + 3 k_s -k_{i,1})k_r + 9k_r^2}.
\end{equation}
However, if one assumes that $k_{i,0} = k_{i,1} = k_i$ this simplifies to
\begin{equation}
Z(t_p) = \alpha e^{-t/\tau_1} + \gamma
\end{equation}
with 
\begin{equation}
\tau_1 = \frac{1}{(k_{i} + 3 k_r)}.
\end{equation}
For recombination processes, an analogue evaluation can be performed, starting from  $\Vec{N}(0) = (0,0,1)$.

\subsection{Data Analysis}

\textbf{Calculation of $\rho$ and $c$}\\
Under orange laser excitation, only populations $M_0$ and $M_1$ contribute to the fluorescence signal since excitation occurs in a charge-state selective manner. Thus, our experimental observables, the time averaged photoluminescence intensities in the signal and reference traces of the ODMR experiment ($I_{\mathrm{sig}}$ and $I_{\mathrm{ref}}$)  are directly proportional to $M_0$ and $M_1$. For $I_{\mathrm{ref}}$, we define 
\begin{equation}
I_{\mathrm{ref}} = \epsilon_0 M_0 + \epsilon_1M_1 + \epsilon_{-1}M_{-1} = \epsilon_0M_0 + 2\epsilon_1M_1
\end{equation}
where we have introduced $\epsilon_i$ ($i = 0, \pm1$) as proportionality factors transferring level populations for a spin state $m_S = i$ to fluorescence counts. For the NV center $\epsilon_0 >> \epsilon_1$. Also, $M_1 = M_{-1}$ and $\epsilon_1 = \epsilon_{-1}.$   The $\pi$-pulse of the ODMR experiment swaps the populations of the $m_S=0$ and $m_s = 1$ levels,  resulting in
\begin{equation}
I_{\mathrm{sig}} = \epsilon_0 M_1 + \epsilon_1 M_0 + \epsilon_{1}M_1 = (\epsilon_0 + \epsilon_1)M_1 + \epsilon_1 M_0 .
\end{equation}
As a proxy for spin state polarization, we define ODMR contrast $c$
\begin{equation}
c = \frac{I_{\mathrm{ref}}-I_{\mathrm{sig}}}{I_{\mathrm{ref}}} = \frac{(\epsilon_0-\epsilon_1)(M_0 - M_1)}{\epsilon_0M_0 + 2\epsilon_1M_1} \approx \frac{\epsilon_0 (M_0 - M_1)}{\epsilon_0M_0} = \frac{M_0-M_1}{M_0}
\end{equation}
where we have set $\epsilon_1$ to zero, since  $\epsilon_0 >> \epsilon_1$. 
To estimate the relative population $\rho$ of the  NV$^-$ charge state  from the experimental data, we use the linear combination 

\begin{equation}
\frac{1}{3}I_{\mathrm{ref}} + \frac{2}{3} I_{\mathrm{sig}} \propto M_0 + M_1 + M_{-1} \propto \rho
\end{equation}

which is independent of $c$ and reference it to the contrast - corrected value of the ODMR signal after green charge state initialization. \\

\noindent
\textbf{Fitting} \\
The time evolutions of the signal and reference traces of an ODMR experiment as a function of ionizing/recombining laser pulse length, $c_{\mathrm{ODMR}}^{\mathrm{ref}}(t)$ and $c_{\mathrm{ODMR}}^{\mathrm{sig}}(t)$,  were fit simultaneously to
\begin{equation}
c_{\mathrm{ODMR}}^{\mathrm{ref}}(t) = \gamma_1 + \alpha_1\exp{-t/\tau_1} 
\end{equation}
\begin{equation}
c_{\mathrm{ODMR}}^{\mathrm{sig}}(t) = \gamma_1 + \gamma_2 + \alpha_2\exp{-t/\tau_1} 
\end{equation}
for the mono-exponential case and
\begin{equation}
c_{\mathrm{ODMR}}^{\mathrm{ref}}(t) = \gamma_1 + \alpha_1\exp{-t/\tau_1} + \beta_1 \exp{-t/\tau_2} 
\end{equation}
\begin{equation}
c_{\mathrm{ODMR}}^{\mathrm{sig}}(t) = \gamma_1 + \gamma_2 + \alpha_2\exp{-t/\tau_1} + \beta_2 \exp{-t/\tau_2} 
\end{equation}
for the bi-exponential case. Fit uncertainties were extracted with a bootstrapping approach.  \\

\newpage 
\section{Additional Results and Data}

\small
\begin{longtable}[H]{p{1.7cm}|p{2.4cm}|p{2.15cm}|p{2.95cm}|p{2.7cm}|p{2cm}} 
\caption{Numerical results for  ODMR contrast and orange ionization rates, before (b) and after (a) UV/blue induced aging with total energy $E$. Numerical values are reported with 95\% confidence intervals. } \\

NV & $c^{520}$ (b) [\%] & $c^{520}$ (a) [\%]  & $k_i^{594}$ (b) [MHz] & $k_i^{594}$ (a) [MHz] & $E$ [mJ]  \\  \hline\hline
1 &  41.6(6) & - & 0.160(7) & -  & - \\  
2 &  39.2(7)  & - & 0.132(8) & - & - \\ 
3 &  26(3)  &  - & 0.15(2) & -  & -  \\
4 &  9.8(7) & -  & 0.25(3)   & - & - \\
5 &  36.6(7) & - &  0.8(1) & - & - \\
6 &  38.6(6)  & - & 0.28(2)  & -  & - \\
7 &  35.1(8) & - &  0.19(2) & - & - \\
8 &  38.7(4)& -  & 0.22(1) & -  & -  \\
\textbf{Blue Set}  \\ \hline
$\triangleleft$ & 39.4(5) & 38.7(6) & 0.30(1)&  0.27(1) & 6625 \\
\textbf{+} &  41.0(5) & 38.7(6) & 0.038(1) & 0.174(5) & 5583 \\ 
$\diamondsuit$ & 35.7(8) & 37(1) & 0.012(2)& 0.035(3) & 3577 \\ 
\textbf{UV Set}   \\ \hline 
$\triangleright$ & 38.1(4) & 34(1) & 0.55(5) & 1.1(2) & 201 \\
\textbf{$\star$} & 40.6(1) & 27(3) &  0.161(8) & 0.7(1) & 373 \\ 
$\times$  & 41.9(8) & 32(2) & 0.026(1) & 0.10(2)  & 938 \\
$\diamond$ & 38(2) & 36(2) & 0.005(1) & 0.026(5) & 136 
\label{NVnumerics}
\end{longtable}
\newpage 

\newpage 
\begin{figure}
\includegraphics[width=1.0\textwidth]{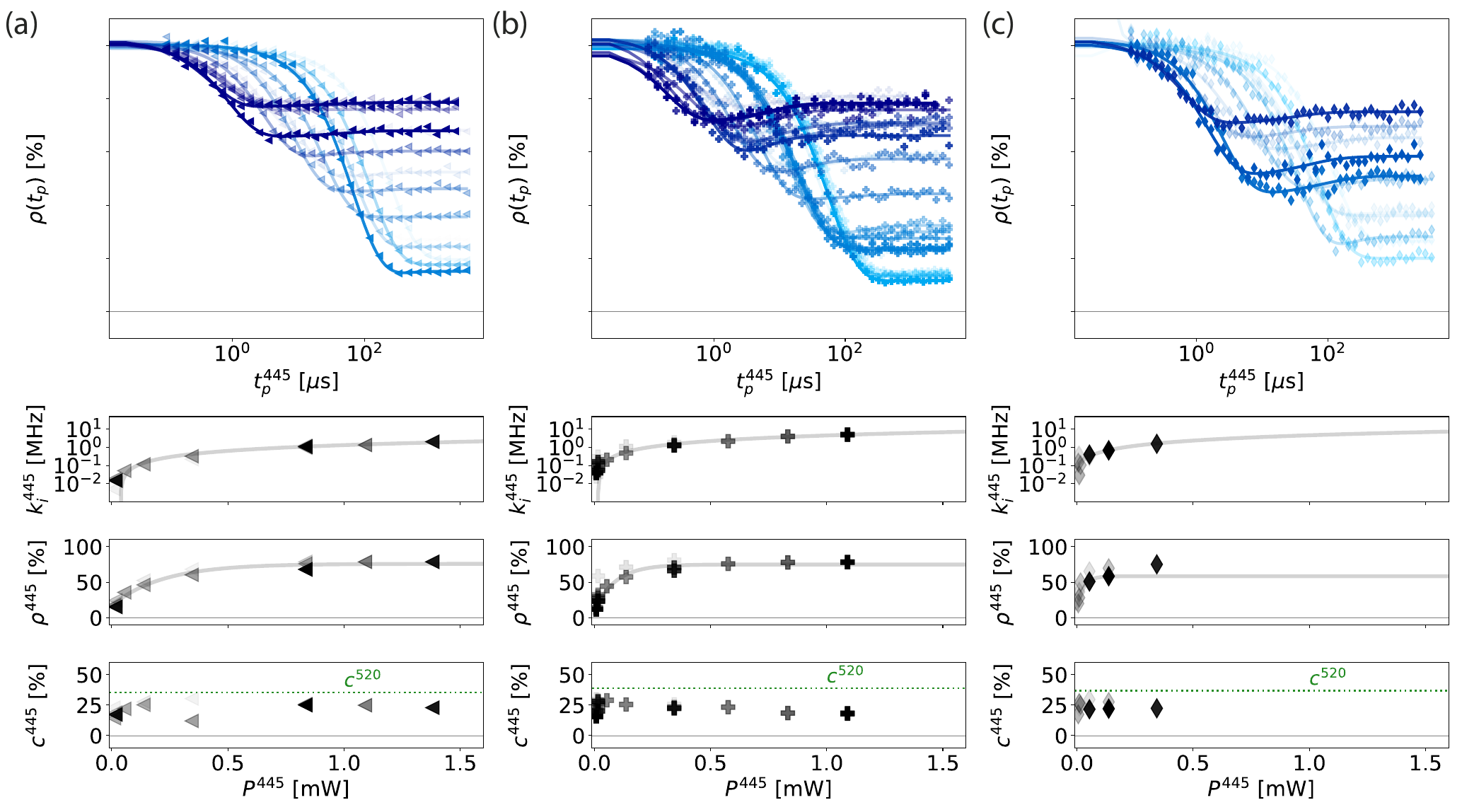}
\caption{Additional data for charge and spin dynamics of NV center under \SI{445}{nm} illumination. For three representative NV centers the following data are shown. \textit{Top}: Evolution of NV$^-$ fraction  as a function of blue pulse lengths. The color of the cure indicates the blue laser power, ranging from $\sim$ \SI{0.1}{mW} to \SI{1}{mW}. The opacity of the curves encode the amount of blue laser light the NV center had already been exposed to when the measurement was taken. Very transparent curves were taken first, on pristine NV centers, while opaque curves show data for 445-aged NV centers. \textit{Bottom}: Ionization rates $k_i^{445}$, steady state $\rho^{445}$ and $c^{445}$ as a function of P$^{445}$; extracted from the curves shown in the top plot. Again, opacity encodes the age of the NV center. For $k_i^{445}$ a linear fit, for steady state $\rho^{445}$ an exponential fit were performed and are in good agreement with the data. For $c^{445}$ the contrast of the NV center under green illumination,$c^{520}$, is indicated with a green dashed line as a reference. }
\label{SIbluepower} 
\end{figure} 
\newpage 

\begin{figure}[ht]
\includegraphics[width=0.5\textwidth]{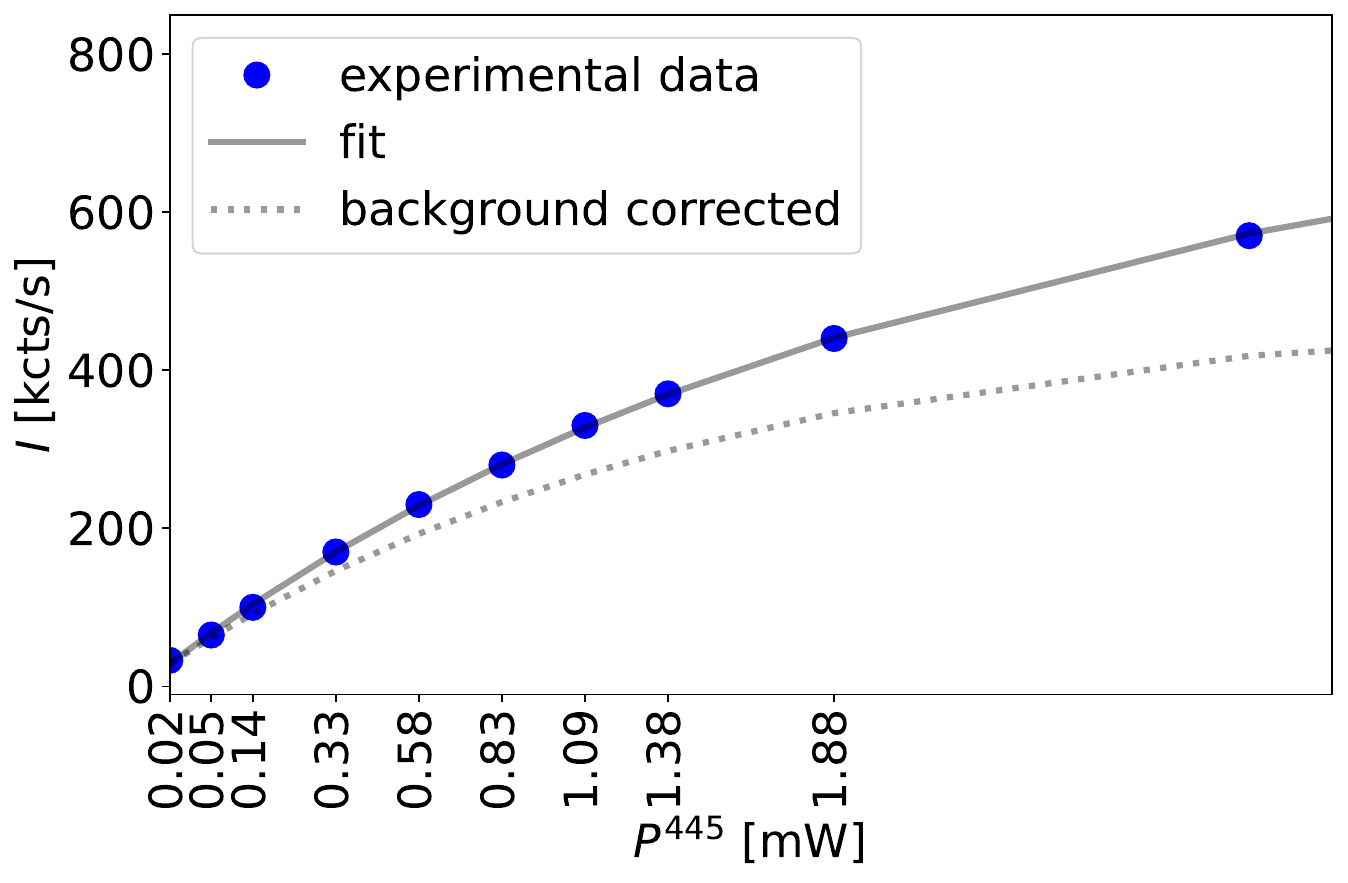}
\caption{Saturation curve of a representative NV center  under \SI{445}{nm} illumination. Since saturation of NV$^{-}$ under \SI{445}{nm} excitation occurs at powers exceeding \SI{1}{mW}, ODMR contrast $c$ is independent of $P^{445}$ in our measurements performed at blue powers $\leq$\SI{1}{mW}. }
\label{SIbluesaturation}
\end{figure}

\begin{figure}[ht]
\includegraphics[width=0.5\textwidth]{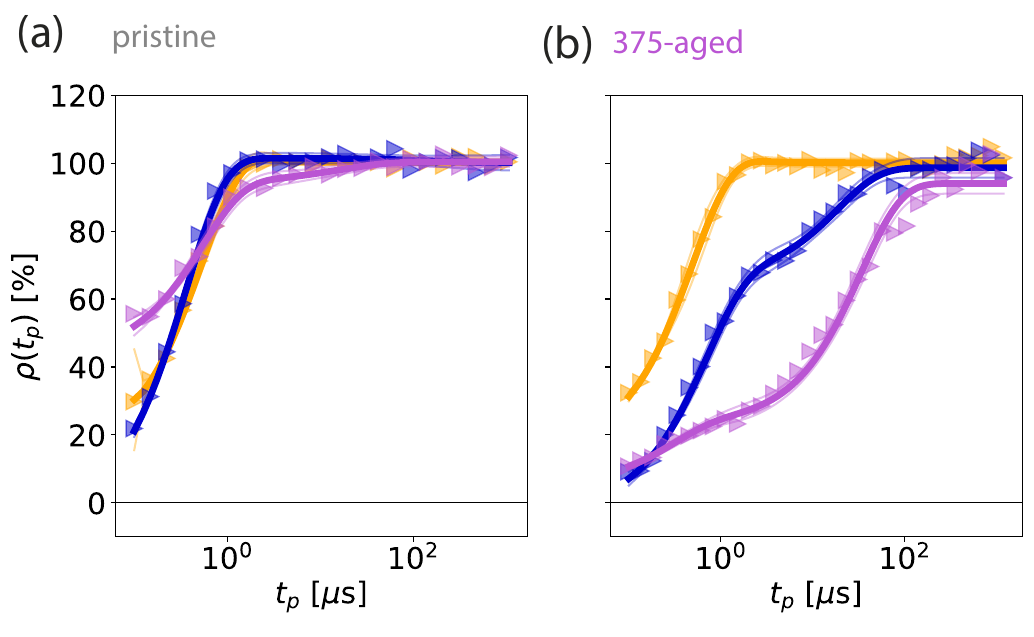}
\caption{Recombination after i) orange, ii) blue and iii) UV induced illumination, measured on the same NV center before (\textbf{a}) and after (\textbf{b}) UV induced aging. The NV center was ionized using 
 sufficiently long ($>$\SI{250}{us}) pulses of the respective color. In each case pulse lengths was chosen based on an ionization measurement. For \SI{594}{nm} and \SI{445}{nm}, ionization was performed at the same laser powers  for both measurements  (\SI{0.010}{mW} for blue). For UV illumination laser powers were \SI{0.069}{mW} and \SI{0.034}{mW} for measurements on the fresh and the aged NV center, respectively. The recombination was performed using green laser pulses with pulse lengths ranging from \SI{100}{ns} -- 
\SI{2}{ms}. }
\label{SItrap}
\end{figure}

\newpage

\begin{figure}[ht]
\includegraphics[width=1.0\textwidth]{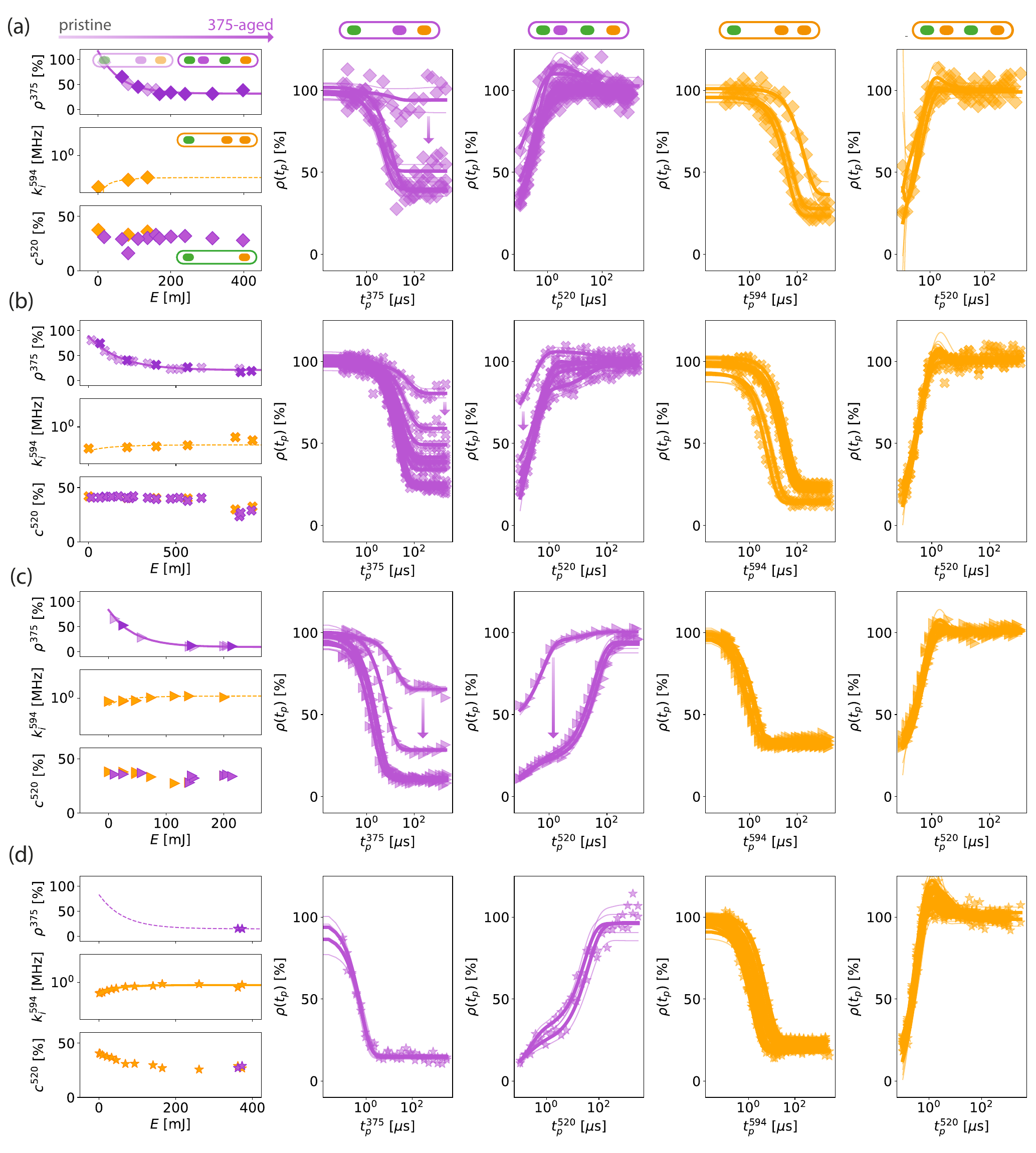}
\caption{Additional data for aging of NV charge environment induced by UV illumination. Each panel a-d shows data for a separate NV center. \textit{Left:} Evolution of $\rho^{375}$, $k_i^{594}$ and $c^{520}$ as a function of exposure dose. For $\rho^{375}$, opaque data points indicate that at this point a recombination measurement was performed; for transparent data points $\rho^{375}$ was determined from the first data point of a recombination measurement.    \textit{From second-left to right:} Ionization under UV/orange laser illumination and recombination after UV/orange induced ionization.}
\label{SIuvage}
\end{figure}

\begin{figure}[ht]
\includegraphics[width=1.0\textwidth]{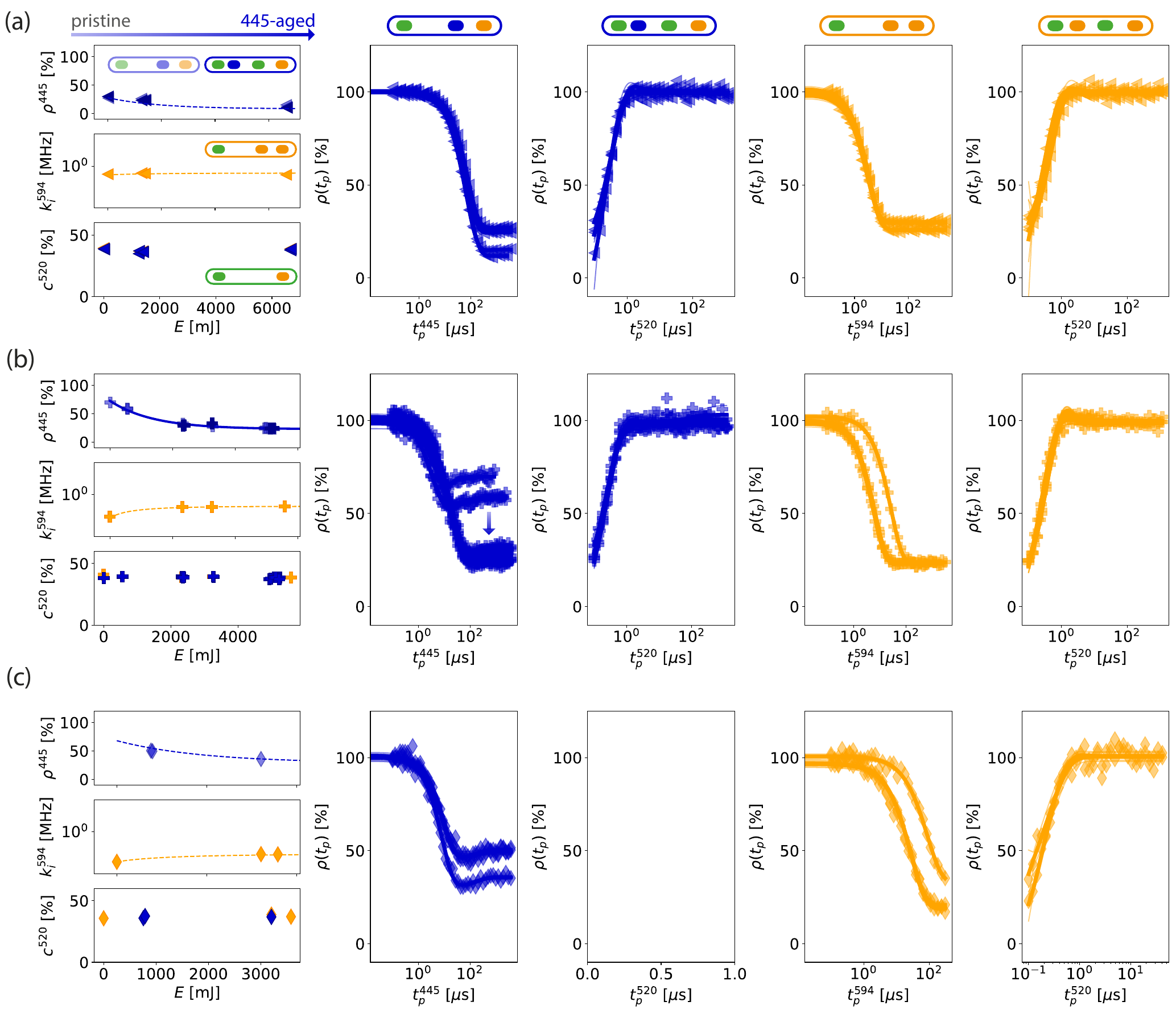}
\caption{Additional data for aging of NV charge environment induced by blue light illumination. Each panel a-c shows data for a separate NV center. \textit{Left:} Evolution of $\rho^{445}$, $k_i^{594}$ and $c^{520}$ as a function of exposure dose. For $\rho^{445}$, opaque data points indicate that at this point a recombination measurement was performed; for transparent data points $\rho^{445}$ was determined from the first data point of a recombination measurement.    \textit{From second-left to right:} Ionization under blue/orange laser illumination and recombination after blue/orange induced ionization.}
\label{SIblueage}
\end{figure}

\FloatBarrier

\bibliography{biblio}